\documentclass[aps,pre,reprint,amsmath,amssymb,floats,longbibliography,superscriptaddress]{revtex4-2}

\usepackage{graphicx}
\usepackage{bm}
\usepackage{newunicodechar}
\usepackage[utf8]{inputenc}
\usepackage[T1]{fontenc}
\usepackage{color}

\begin{document}
\title{Nonequilibrium phase transitions in a Brownian $p$-state clock model}
\author{Chul-Ung Woo}
\affiliation{Department of Physics, University of Seoul, Seoul 02504, Korea}
\author{Jae Dong Noh}
\affiliation{Department of Physics, University of Seoul, Seoul 02504, Korea}
\date{\today}

\begin{abstract}
We introduce a Brownian $p$-state clock model in two dimensions and investigate the nature of phase transitions numerically. As a nonequilibrium extension of the equilibrium lattice model,  the Brownian $p$-state clock model allows spins to diffuse randomly in the two-dimensional space of area $L^2$ under periodic boundary conditions. We find three distinct phases for $p>4$: a disordered paramagnetic phase, a quasi-long-range-ordered critical phase, and an ordered ferromagnetic phase. In the intermediate critical phase, the magnetization order parameter follows a power law scaling $m \sim L^{-\tilde{\beta}}$, where the finite-size scaling exponent $\tilde{\beta}$ varies continuously. These critical behaviors are reminiscent of the double Berezinskii-Kosterlitz-Thouless~(BKT) transition picture of the equilibrium system. At the transition to the disordered phase, the exponent takes the universal value $\tilde\beta = 1/8$ which coincides with that of the equilibrium system. This result indicates that the BKT transition driven by the unbinding of topological excitations is robust against the particle diffusion. On the contrary, the exponent at the symmetry-breaking transition to the ordered phase deviates from the universal value $\tilde{\beta} = 2/p^2$ of the equilibrium system. The deviation is attributed to a nonequilibrium effect from the particle diffusion.
\end{abstract}

\maketitle

\section{Introduction}\label{sec:intro}
Phase transitions and critical phenomena have been extensively studied in thermal equilibrium systems. Recently, there has been renewed interest in these topics due to the observation of diverse and complex collective behaviors in nonequilibrium systems. In particular, there has been a growing interest in nonequilibrium active matter systems composed of self-propelled particles. Active matter systems are ubiquitous in nature, spanning from biological systems such as swarming bacteria and migrating birds to synthetic material systems such as Janus particles~\cite{marchetti2013hydrodynamics,elgeti2015physics,bechinger2016active}.
Particle motility gives rise to a distinct collective phenomenon that cannot be observed in equilibrium systems~\cite{cates2015motility,di2010bacterial,cavagna2014bird,ballerini2008interaction,ward2008quorum,cavagna2017dynamic,kumar2014flocking}. The Vicsek model is a prototypical example exhibiting this property~\cite{vicsek1995novel}. It is comprised of self-propelled particles equipped with a local velocity-alignment interaction in two dimensions. Despite having a short-ranged interaction with continuous rotational symmetry, the Vicsek model demonstrates a long-range ordered phase, which is prohibited in equilibrium systems due to the Mermin-Wagner theorem~\cite{mermin1966absence}.

In contrast to particles in equilibrium lattice models, self-propelled particles exhibit persistent motion. In particular, persistence makes them different from passive Brownian particles. While the effects of mobility and persistence on collective behaviors have been extensively studied, the exclusive influence of mobility has received little attention. Recently, we have investigated the nature of phase transitions in the Brownian $q$-state Potts model in two dimensions~\cite{woo2022suppression}. In this model, Potts spins perform a passive Brownian motion and exchange a ferromagnetic interaction with local neighbors. The passive Brownian motion disturbs the propagation of the spin-spin correlations, which prohibits a phase coexistence. As a result, the Brownian Potts model undergoes a continuous phase transition even for $q>4$~\cite{woo2022suppression}. This study motivates us to investigate the phase transition in the Brownian $p$-state clock model, in which clock spins are allowed to move freely in the two dimensional space.

The equilibrium $p$-state clock model on the two-dimensional lattice~\cite{Ortiz.2012,Li.2020} is defined by the Hamiltonian
\begin{equation}\label{Heq}
    \mathcal{H}_{\text {eq}}= -J \sum_{\langle i, j \rangle} \cos(\theta_i - \theta_j),
\end{equation}
where $\theta_i = 2\pi k_i / p$ is the clock spin variable with $k_i=0,1\cdots, p-1$ at lattice site $\bm{r}_i$, $J>0$ is a ferromagnetic interaction strength, and $\langle \cdots\rangle$ denotes a pair of nearest neighbor sites. The model reduces to the $XY$ model with the continuous rotation symmetry in the limit $p\to\infty$~\cite{Kosterlitz.1973}. On the symmetry basis, the $p$-state clock model can be regarded as the $XY$ model perturbed with a potential $-h_p \sum_i \cos p\theta_i$ of discrete $p$-fold symmetry~\cite{jose1977renormalization}. 

The equilibrium model has been investigated thoroughly using the renormalization group~(RG) theory~\cite{jose1977renormalization}, which predicts a high temperature disordered phase, an intermediate quasi-long-range-ordered~(QLRO) phase, and a low temperature ordered phase separated by the Berezinskii-Kosterlitz-Thouless~(BKT) transitions for $p>4$~\cite{berezinskii.1971, Kosterlitz.1973, Izyumov.1988, jose1977renormalization}. In the QLRO phase, spin wave excitation results in a power law decay of the spin-spin correlation function, $\left\langle e^{i(\theta_i - \theta_j)} \right\rangle \sim |\bm{r}_i -\bm{r}_j|^{-(d-2+\eta)}$ with the correlation exponent $\eta$ varying continuously within the range 
\begin{equation}\label{eta_QLRO}
\frac{4}{p^2} \le \eta \le \frac{1}{4}. 
\end{equation}
There also exists a topological excitation, vortices and antivortices. They form bound pairs in the QLRO phase. It is the unbinding of the vortex-antivortex pairs that drives the transition to the disordered phase, known as the BKT transition~\cite{berezinskii.1971, Kosterlitz.1973}. At the transition, $\eta$ takes the universal value $1/4$. The $p$-fold clock symmetry is broken spontaneously in the ordered phase. The symmetry-breaking transition also belongs to the BKT transition universality class, at which $\eta=4/p^2$~\cite{jose1977renormalization}. The RG scenario has been confirmed in numerical Monte Carlo simulation studies~\cite{Noh.2002, tobochnik1982properties, lapilli2006universality, tuan2022binder}. 

We generalize the equilibrium $p$-state clock model on a lattice by allowing the clock spins to diffuse freely in the continuous two-dimensional space of area $L^2$. In Sec.~\ref{sec:model}, we introduce the model and explain the time evolution rule. 
With extensive Monte Carlo simulations, we investigate the phase diagram and the critical behavior. We have performed the Monte Carlo simulations at several values of $p\geq 4$ for the phase diagram. We have investigated the critical behavior in greater detail at $p=8$ with a focus on similarities and differences between the equilibrium clock model on a lattice and the Brownian clock model. The main numerical results will be presented in Sec.~\ref{sec:numerical}. 
Finally, we conclude the paper with a summary and discussions in Sec.~\ref{sec:summary}.

\section{Brownian $p$-state clock model}\label{sec:model}

The Brownian $p$-state clock model is composed of $N=\rho L^2$ particles of
density $\rho$ diffusing in the two-dimensional continuous space of area
$L\times L$ under periodic boundary conditions. A particle $i$ carries a
clock spin $\bm{u}_i = (\cos\theta_i, \sin\theta_i)$ with $\theta_i \in \{0,
2\pi/p, \cdots, 2(p-1)\pi/p\}$. Particles interact ferromagnetically with the others within a unit distance $r_0=1$. The spin-spin interaction is represented with a Hamiltonian 
\begin{equation}\label{Hbcm}
        \mathcal{H}[\{\bm{r}_i\}, \{\theta_i\}] 
          = -K \sum_{ |\bm{r}_i - \bm{r}_j| < r_0} \cos{\left(\theta_i-\theta_j\right)},
\end{equation} 
where $\bm{r}_i$ denotes the position of particle $i$ and $K>0$ represents the ferromagnetic interaction strength.

Positions and spins are updated at discrete time steps according to the following rule: (i) A particle performs a jump of length $l_0$ in a random direction. (ii) A particle $i$ attempts a spin flip $\theta_i \to \theta_i + \Delta\theta$, which is accepted with the probability $p_{\rm acc.} = \min\left[1, \exp(-\Delta_i \mathcal{H})\right]$ where $\Delta_i \mathcal{H}$ is the change in the Hamiltonian under a spin update $\theta_i\to\theta_i+\Delta\theta$ while keeping all the other spins fixed. For $p \le 8$, we choose $\Delta \theta$ among $\{0, \pm 2\pi/p\}$ randomly and independently. When $p\ > 8$, $\Delta \theta$ is chosen among multiples of $2\pi/p$ within the interval 
$[-\pi/4, \pi/4]$. As in Ref.~\cite{woo2022suppression}, we adopt parallel update dynamics.

The spin dynamics obeys the local detailed balance with respect to the
Hamiltonian $\mathcal{H}$. The particle diffusion, however, breaks the
detailed balance.
The Brownian clock model could be regarded as being in thermal contact with
two distinct heat baths, each coupled to the spin and position degrees of
freedom, respectively. Those heat baths will be referred to as the spin bath
and the diffusion bath, respectively. We can generalize the model by
adopting a stochastic particle hopping dynamics with the Metropolis rule
with the probability $p_h = \min\left[1, e^{-a\Delta_{h}
\mathcal{H}}\right]$ accepting a particle hopping. Here, $\Delta_h
\mathcal{H}$ denotes the change in the Hamiltonian~\eqref{Hbcm} upon a
hopping. Then, the
parameter $a$ corresponds to the ratio of the spin bath
temperature and the diffusion bath temperature. 
{ In this study with $a=0$, particle hoppings are always accepted as if the system were
in contact with an infinite temperature diffusion bath.}
Thus, the Brownian clock model is driven out of
equilibrium. As a nonequilibrium effect, we will examine the energy flow
between the two baths through the Brownian clock spin system in
Sec.~\ref{Heat flow}.
{
When $a=1$, the dynamics satisfies the detailed balance condition and the model
corresponds to an equilibrium annealed clock model. An equilibrium annealed diluted XY model in two-dimensional lattices has been studied in 
Ref.~\cite{Popov.2019}, which demonstrates intriguing dynamical behaviors
caused by particle hoppings. The equilibrium limit of the Brownian clock
model~($a=1$) is out of the scope and will not be studied in this work.}

When $l_0=0$, particles do not move and the Brownian clock model reduces to
the equilibrium clock model with quenched disorder in the particle position.
In the opposite limiting case with $l_0=\infty$, we expect that the model
can be described by the mean field theory. In this work, we focus on the
case with finite nonzero $l_0$.

It is worthy to mention the difference between the Brownian clock model and
the other {\em active} spin models such as the active Ising
model~\cite{solon2013revisiting,solon2015phase,solon2015flocking}, the
active Potts model~\cite{chatterjee2020flocking,mangeat2020flocking}, and
the active clock model~\cite{chatterjee2022polar,solon2022susceptibility}.
In the latter, spins interact ferromagnetically and point toward the
direction of self-propulsion. On the contrary, the Brownian clock model
consists of {\em passive} particles. They move independently and randomly
irrespective of their spin state. Our study aims at filling a gap between
ordering phenomena of spins frozen in lattices and of self-propelled active
particles. We also expect that a Brownian spin model can be applied to
investigate opinion dynamics in complex systems consisting of mobile
agents~\cite{Ferri.2023} and can be a starting point for the study of
multi-species active particle systems~\cite{Menzel.2012,Chatterjee.2023}.

We have performed Monte Carlo simulations at various values of $p$ with fixed interaction range~($r_0=1$) and the hopping length~($l_0=0.5$). Varying the particle density $\rho$ and the coupling strength $K$, we have measured the magnetization 
\begin{equation}
    m\equiv \left\langle \left|\frac{1}{N}\sum_{i=1}^N \bm{u}_i \right|\right\rangle,
\end{equation}
as an order parameter. The angular bracket $\langle \cdot \rangle$ denotes a time average in the steady state.
The maximum system size is $L=512$ at which simulations were run up to $4\times 10^8$ time steps.

\begin{figure}[t]
    \includegraphics*[width=\columnwidth]{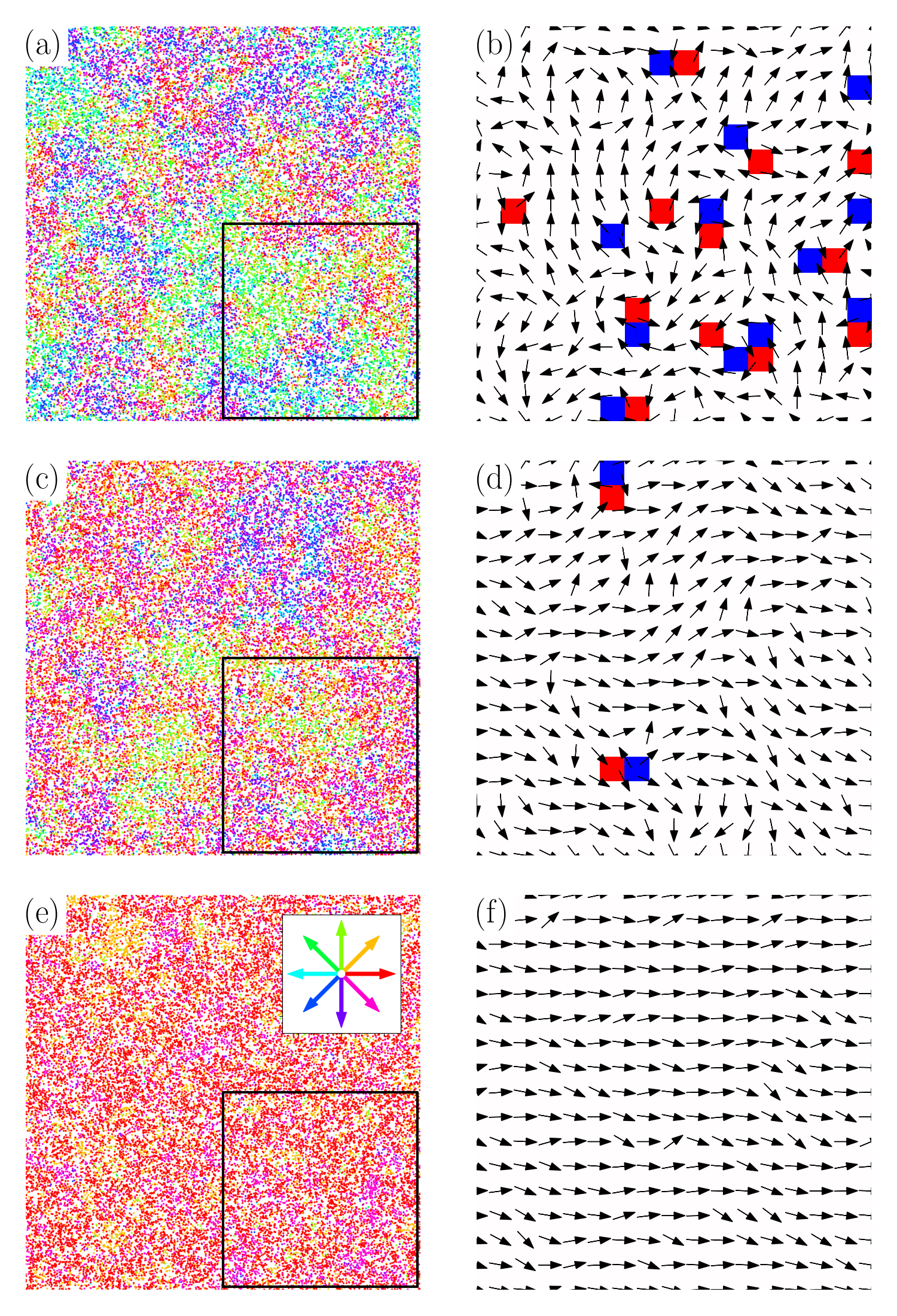}
    \caption{Typical configurations of the Brownian clock model with $p=8$, $\rho=2$, and $L=128$ in the disordered phase [(a) and (b) with $K=0.5$], the QLRO phase [(c) and (d) with $K=0.6$], and the ordered phase [(e) and (f) with $K=1.5$]. In (a,c,e), each dot represents a particle whose spin state is color-coded according to the chart in (e). Spin configurations inside the square boxes are coarse-grained and shown in (b,d,f), where each arrow represents an average direction of spins in a $4\times 4$ square cell. Blue and red squares in (b) and (d) represent vortices and antivortices, respectively, which will be defined in Sec.~\ref{KTphase}.
    }
    \label{fig1}
\end{figure}

\begin{figure}[t]
    \includegraphics[width=\columnwidth]{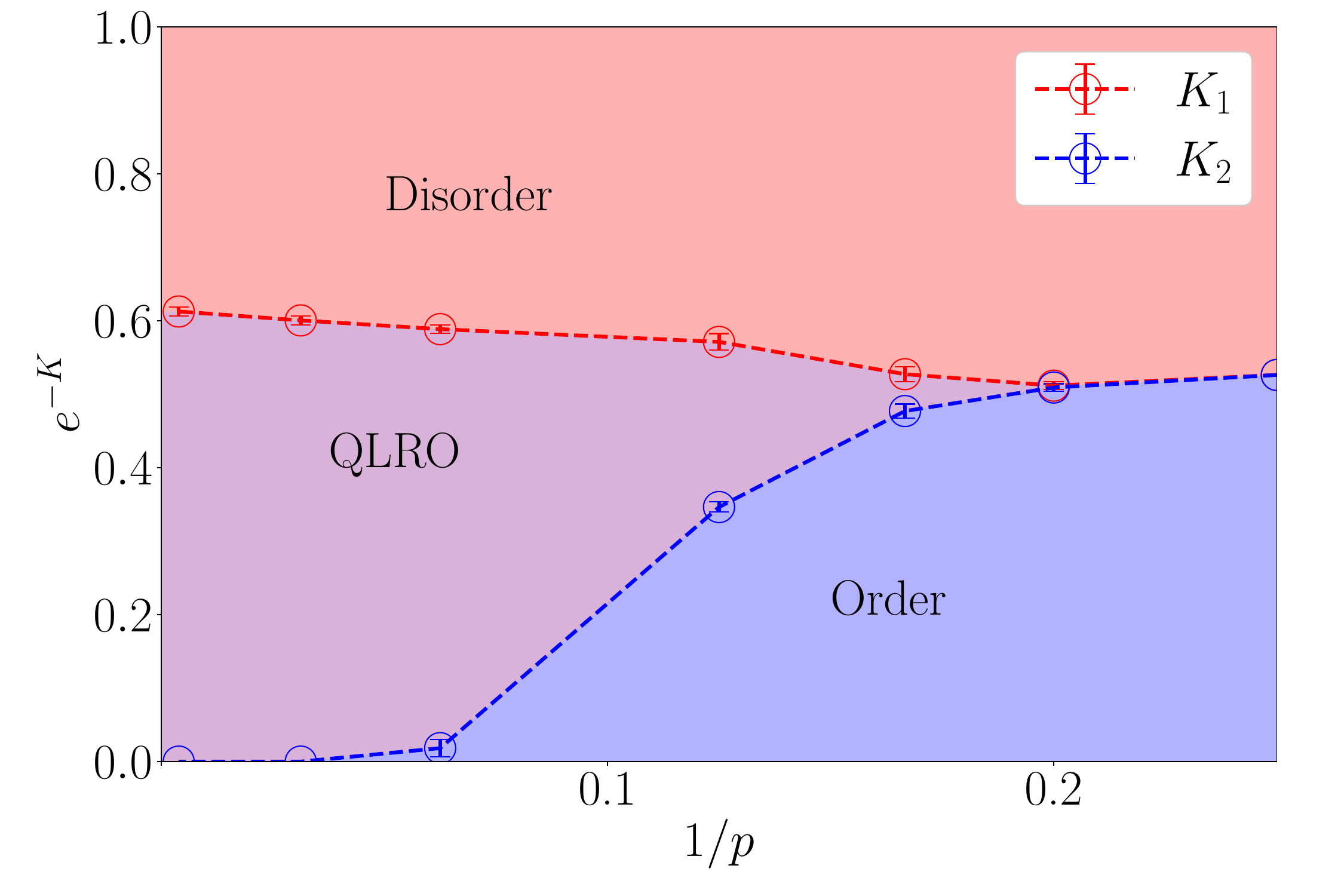}
    \caption{Phase diagram of the Brownian $p$-state clock model with fixed particle density $\rho=2$. Lines are guides to the eyes.}
    \label{fig2}
\end{figure}

Typical configurations at three different values of $K$ at $p=8$ and
$\rho=2$ are shown in Fig.~\ref{fig1}. A phase diagram can be obtained from
the finite-size-scaling analysis of the order parameter. Figure~\ref{fig2}
shows a phase diagram at $\rho=2$, which consists of three distinct phases:
a disordered phase, a QLRO phase, and an ordered phase. It has the same
overall structure as the phase diagram of the equilibrium system. Details of
the phase diagram and the nature of the phase transitions will be presented
in the following section. 

\section{Numerical results}\label{sec:numerical}

In this section, we present the numerical results for the critical behavior of the Brownian $p$-state clock model. We have investigated the critical behavior at several values of $p$ and obtained the similar results. Thus, we only present the results at $p=8$.

\subsection{Finite size scaling of the order parameter}\label{FSS}
\begin{figure}[t]
    \includegraphics[width=1\linewidth]{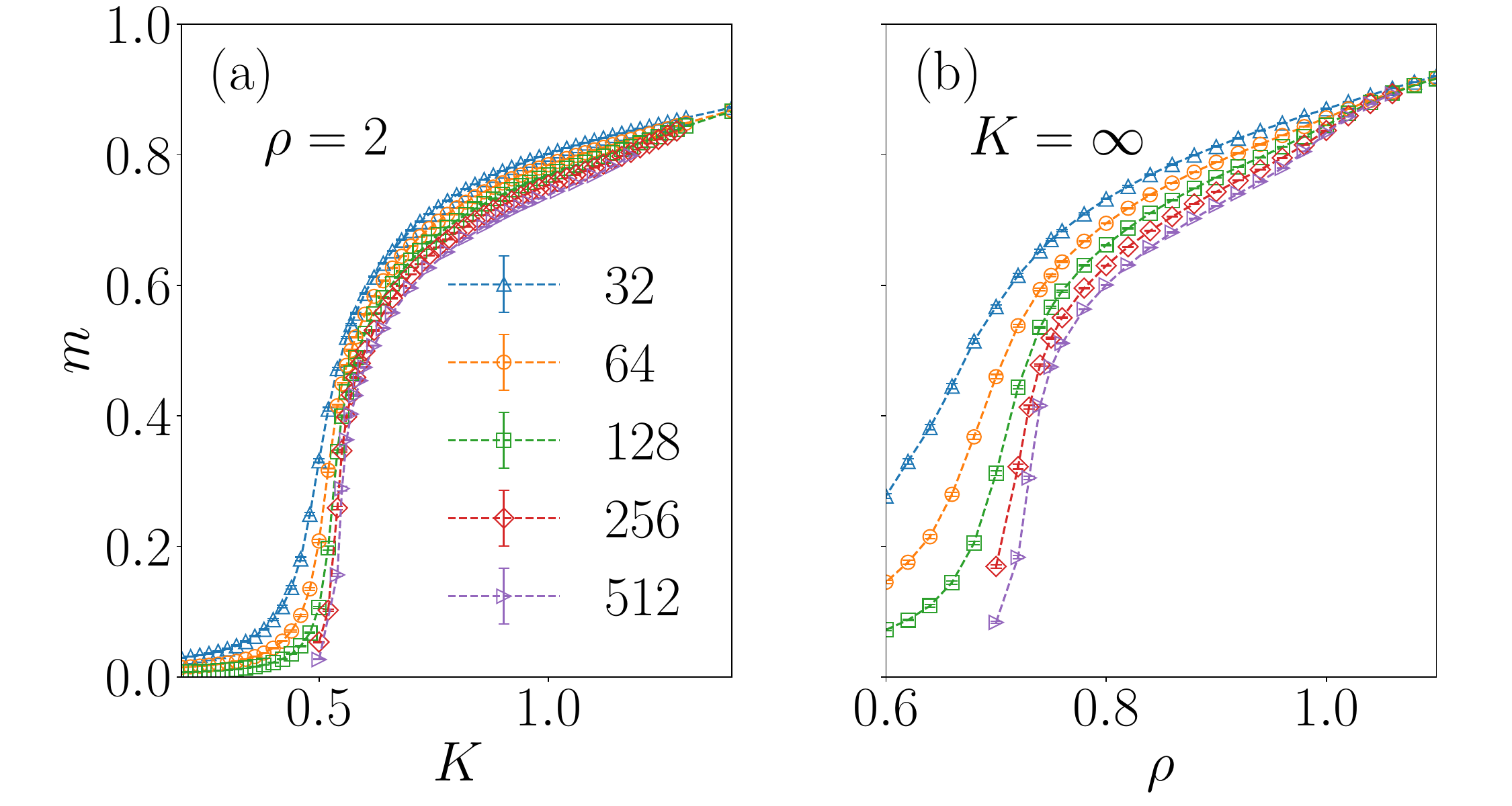}
    \caption{
    The order parameter $m$ of the Brownian $8$-state clock model as a function of the coupling constant $K$ with fixed $\rho=2$ in (a) and the particle density $\rho$ with fixed $K=\infty$ in (b) at several values of $32 \le L \le 512$. 
    }
    \label{fig3}
\end{figure}

Figure~\ref{fig3} shows the order parameter $m(K, \rho; L)$ as a function of  $K$ or $\rho$. When $K$ or $\rho$ is large, the order parameter converges to a finite value as $L$ increases. It signals a long-range ordered phase where the $p$-fold clock symmetry is broken spontaneously. In the opposite case, the order parameter decays to zero as $L$ increases. 

Interestingly, the system undergoes phase transitions even when the ferromagnetic spin interaction is infinitely strong~($K=\infty$). Without particle diffusion, it corresponds to the trivial zero temperature limit.  
In the Brownian clock model, however, the particle diffusion generates a temporal fluctuation: It may turn on and off the interaction between particles. It also mixes particles from different magnetic domains. Consequently, the spin ordering cannot be perfect and the system can undergo phase transitions as one varies, e.g., the particle density.

At criticality, the order parameter follows a power-law scaling
\begin{equation}
    m \sim L^{-\tilde{\beta}}
\end{equation}
with the finite-size-scaling exponent denoted as $\tilde\beta$.
We can characterize the finite size scaling behavior of the order parameter and locate the critical point using an effective exponent  
\begin{equation}
    \tilde{\beta}_{\rm eff}(K,\rho;L) \equiv - \frac{\ln [m(K,\rho;2L)]-\ln
    [m(K,\rho;L)]} {\ln 2}.    
\end{equation}
It will converge to the scaling exponent $\tilde\beta$ as $L$ increases in the critical phase. On the other hand, it will converge to $0$ in the ordered phase~($m=O(L^0)$) and to $1$ in the disordered phase~($m = O(L^{-1})$).

\begin{figure}[t]
    \includegraphics[width=1\linewidth]{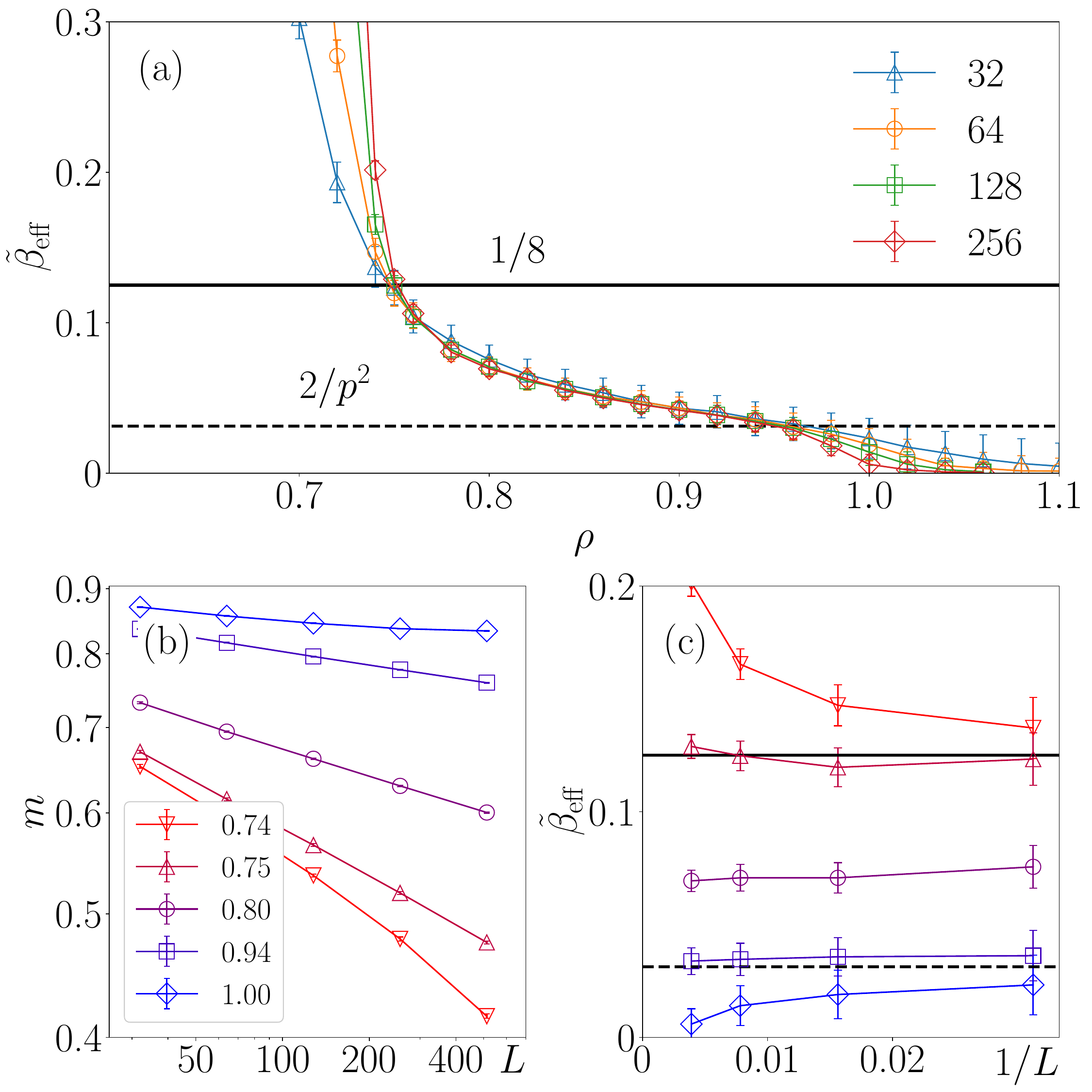}
    \caption{
    Finite size scaling analysis for the order parameter at $K=\infty$.
    (a) Effective exponent $\tilde\beta_{\rm eff}$ vs $\rho$ at several values of $L$. The effective exponent converges to a nontrivial value $\tilde\beta$ when $\rho_1 \le \rho \le \rho_2$ with $\rho_1 \simeq 0.76$ and $\rho_2 \simeq 0.94$. Its value is in agreement with $1/8$~(solid line) at $\rho=\rho_1$ and with $2/p^2=1/32$~(dashed line) at $\rho=\rho_2$.
    In (b) and (c), we present the $L$ dependence of the order parameter and the effective exponent at selected values of $\rho$. The solid and dashed lines in (c) represent the reference values $1/8$ and $1/32$, respectively.
    }
    \label{fig4}
\end{figure}
The effective exponent for the data in Fig.~\ref{fig3}(b) is presented in Fig.~\ref{fig4}.
We can identify three distinct phases: (i) When $\rho < \rho_1 (\simeq
0.76)$, the effective exponent tends to 1 as $L$ increases. This region
corresponds to the disordered phase. (ii) When $\rho_1 < \rho < \rho_2
(\simeq 0.94)$, the effective exponent converges to a nontrivial value
$\tilde{\beta}$ that varies continuously with $\rho$. This region
corresponds to the critical QLRO phase. (iii) When $\rho > \rho_2$, the
effective exponent converges to $0$. This region corresponds to the ordered
phase. 
When the particle density is low, a dynamic interaction
    network, composed of edges between particles whose distance is smaller
    than $r_0$, cannot form an infinite percolation
    cluster~\cite{woo2022suppression}. Thus, the
    particle density should be larger than a threshold value for the ordered
and the QLRO phases even with $K=\infty$.

The finite size scaling behavior of the Brownian clock model reminds us of the double BKT transition picture of the equilibrium clock model on the lattice. As introduced in Sec.~\ref{sec:intro}, the equilibrium $p$-state clock model~($p>4$) has the intermediate QLRO critical phase which is separated from the disordered phase and the ordered phase with the BKT transitions~\cite{jose1977renormalization}. In the QLRO phase, the spin-spin correlation function decays algebraically as $C(r) \sim r^{-(d-2+\eta)}$ with the continuously varying exponent $\eta$ within the range shown in Eq.~\eqref{eta_QLRO}. The power-law decay of the correlation function implies that the order parameter follows the power-law scaling $m \sim L^{-\tilde{\beta}}$ with $\tilde{\beta}= (d-2+\eta)/2$. With $d=2$, the finite-size-scaling exponent also  varies continuously in the range
\begin{equation}\label{beta_QLRO}
    \frac{2}{p^2} \le \tilde{\beta} \le \frac{1}{8} .
\end{equation}

The Brownian clock model also has the QLRO phase with the continuously varying finite-size-scaling exponent $\tilde\beta$. Furthermore, in Fig.~\ref{fig4}, we find that $\tilde{\beta}$ is close to the universal value $1/8$ at the transition point to the disordered phase and $2/p^2=1/32$ at the transition point to the ordered phase. This agreement suggests that the BKT transition may be robust against the particle diffusion.

\begin{figure}[t]
    \includegraphics[width=1\linewidth]{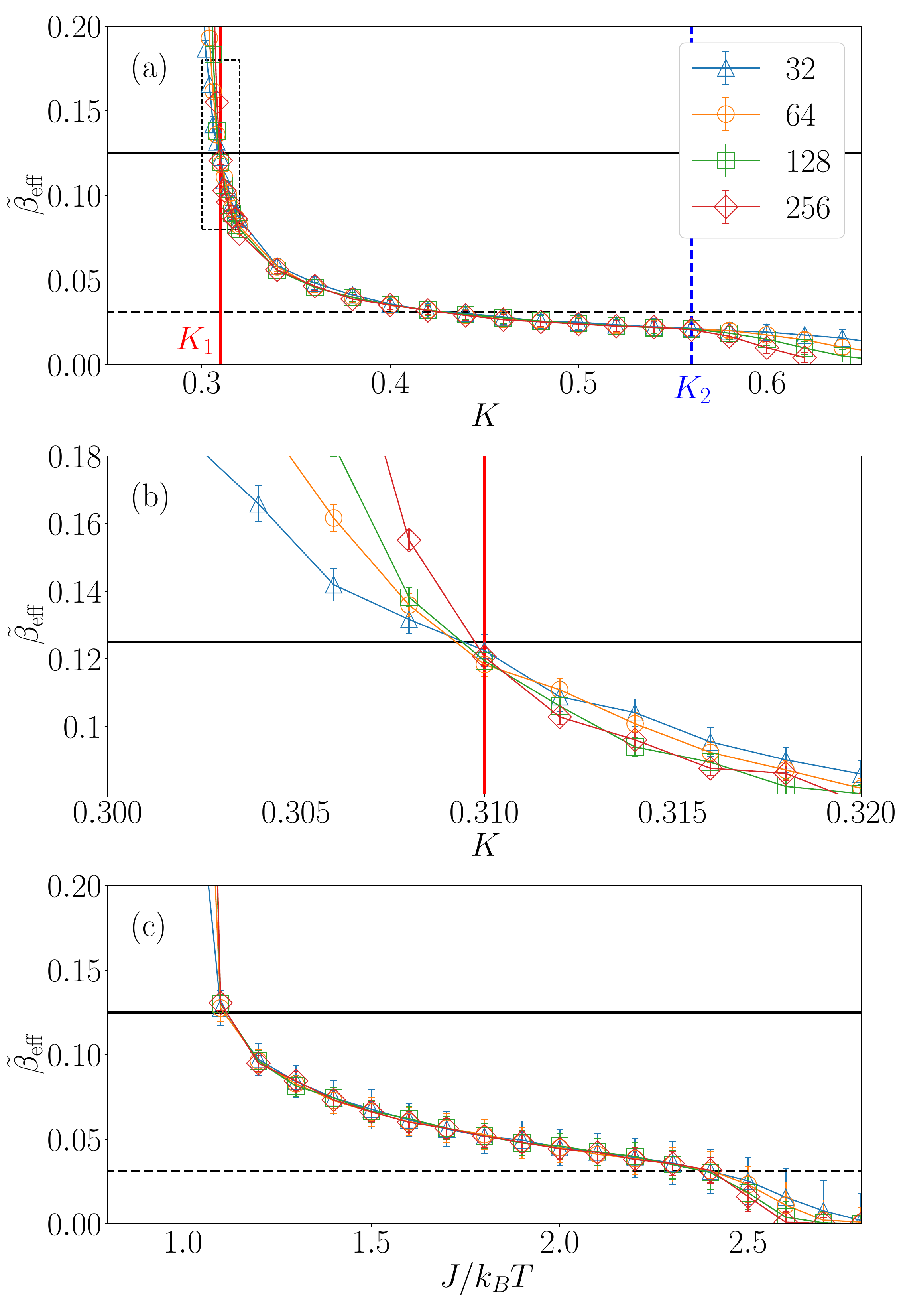}
    \caption{Finite size scaling analysis of the order parameter at $\rho=3$.
    (a)~The effective exponent $\tilde\beta_{\rm eff}$ is plotted as a function of $K$. The vertical lines at $K_1$~(solid, red) and $K_2$~(dotted, blue) indicate the phase transition points. Also drawn are the horizontal lines at $\tilde\beta_{\rm eff}= 1/8$~(solid) and $1/32$~(dashed), which represent the universal values at the BKT transitions of the equilibrium system. 
    (b) Enlarged plot for the region inside the dotted box in (a).
    (c) $\tilde\beta_{\rm eff}$ of the equilibrium model on the lattice as a function of $J/k_B T$. 
    }
    \label{fig5}
\end{figure}

We have also performed the finite-size-scaling analysis at other parameter values. Figure~\ref{fig5} (a) presents the effective exponent plot as a function of $K$ at fixed  $\rho=3$. There are three regions, which correspond to the disordered phase~($K<K_1$), the QLRO phase~($K_1 \le K \le K_2)$, and the ordered phase~($K>K_2$), respectively. In the QLRO phase, the effective exponent converges to a finite-size-scaling exponent $\tilde\beta$ whose value varies with $K$. At the transition point to the disordered phase~($K=K_1$), the exponent is in agreement with the universal value $1/8$ of the BKT transition~(see Fig.~\ref{fig5}(b)). On the other hand, at the symmetry-breaking transition~($K=K_2)$, its value is significantly smaller than the universal value $2/p^2=1/32$. 
This is in sharp contrast to the equilibrium model on the lattice. In Fig.~\ref{fig5}(c), we present the effective exponent plot for the equilibrium model with the Hamiltonian~\eqref{Heq}. The finite-size-scaling exponent takes the universal values $1/8$ and $2/p^2=1/32$ at each BKT transition point. 

\begin{figure}    
\includegraphics[width=1\linewidth]{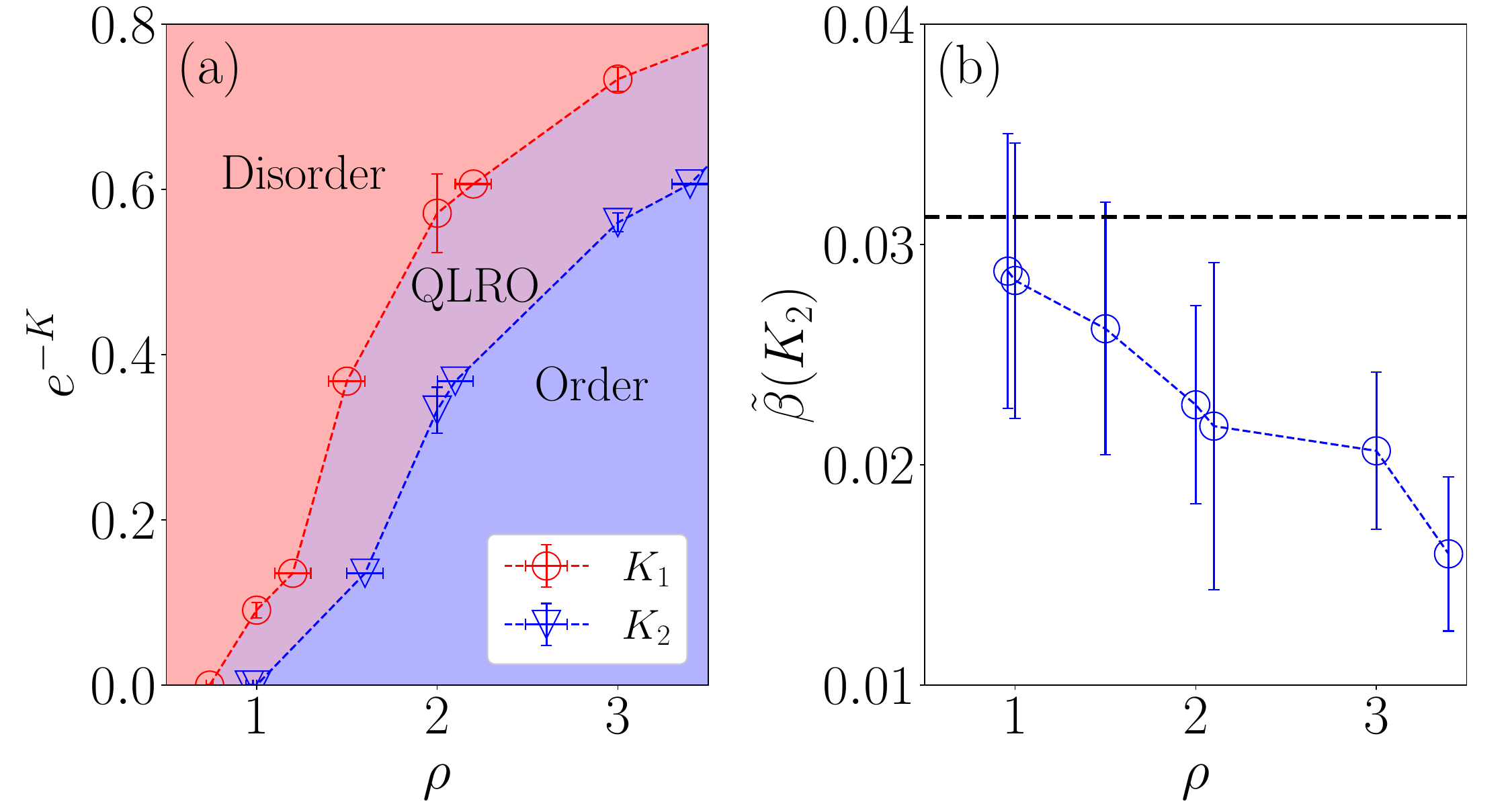}
\caption{(a) The phase diagram of the Brownian eight-state clock model in the $(\rho, e^{-K})$ plane. (b) shows the finite-size-scaling exponent $\tilde\beta$ evaluated at the transition points $K=K_2(\rho)$ between the QLRO phase and the ordered phase. Deviation from the universal value $1/32$ of the equilibrium model becomes large as $\rho$ increases.}
\label{fig6}
\end{figure}
Repeating the finite-size-scaling analysis, we obtain the phase diagram of the Brownian $(p=8)$-state clock model in the $(\rho,e^{-K})$ plane, which is shown in Fig.~\ref{fig6}(a). Along the boundary between the disordered and the QLRO phases, $\tilde\beta$ is in agreement with the universal value $1/8$ of the BKT transition up to a numerical accuracy. On the other hand, along the boundary between the QLRO and the ordered phase, $\tilde\beta$ deviates from the universal value $2/p^2$ of the equilibrium system~(see Fig.~\ref{fig6}(b)). The discrepancy is minimum at $K=\infty$, and becomes large as $\rho$ increases.
Both the coincidence with and the discrepancy from the equilibrium theory at
either boundary of the QLRO phase are puzzling. We will address this issue
in the upcoming sections.

\subsection{BKT phase transition between the QLRO phase and the disordered phase}\label{KTphase}
In the equilibrium model, the BKT transition from the QLRO phase to the disordered phase is driven by the unbinding of vortex-antivortex pairs. These topological excitations are also present in the Brownian clock model. In order to identify them, we divide the two-dimensional plane into a regular array of $(L/b)^2$ cells of size $b^2$ and assign a cell spin $\bm{s}_\alpha = \sum_{i\in \alpha} \bm{u}_i$ with azimuthal angle $\theta_\alpha = \text{arg}\bm{s}_\alpha$ to each cell $\alpha=1,\cdots, (L/b)^2$. These cells form a square lattice. The vorticity can be defined at each elementary square. Let $(\bm{s}_{1}, \bm{s}_{2}, \bm{s}_{3}, \bm{s}_{4})$ be four cell spins enclosing an elementary square in the counterclockwise direction. The vorticity at an elementary square is defined by the winding number given by 
\begin{equation}\label{vorticity}
    v = \frac{1}{2\pi} (\Delta\theta_{12} + \Delta\theta_{23}+\Delta\theta_{34} + \Delta\theta_{41} ),
\end{equation}
where the angle difference $\Delta \theta_{\alpha\alpha'} \equiv \theta_{\alpha'}-\theta_\alpha$ is measured in the range $-\pi < \Delta\theta_{\alpha\alpha'} \le \pi$~\cite{tobochnik1979monte, loft1987numerical} . A nonzero value of $v$ indicates a vortex~($v=+1$) or an antivortex~($v=-1$) core. 

Figures~\ref{fig1} (b), (d), and (f) display a typical cell spin configuration in each phase. We also depict the vortex and antivortex cores identified with the vorticity in Eq.~\eqref{vorticity}.
In the QLRO phase, there are only a few vortices and antivortices, which form bound pairs~(see Fig.~\ref{fig1}(d)). The disordered phase is more abundant in the topological excitation. More importantly, there appear free vortices and antivortices. This qualitative feature supports that the transition from the QLRO phase to the disordered phase belongs to the BKT transition universality.

\begin{figure}[t]
    \includegraphics[width=1\linewidth]{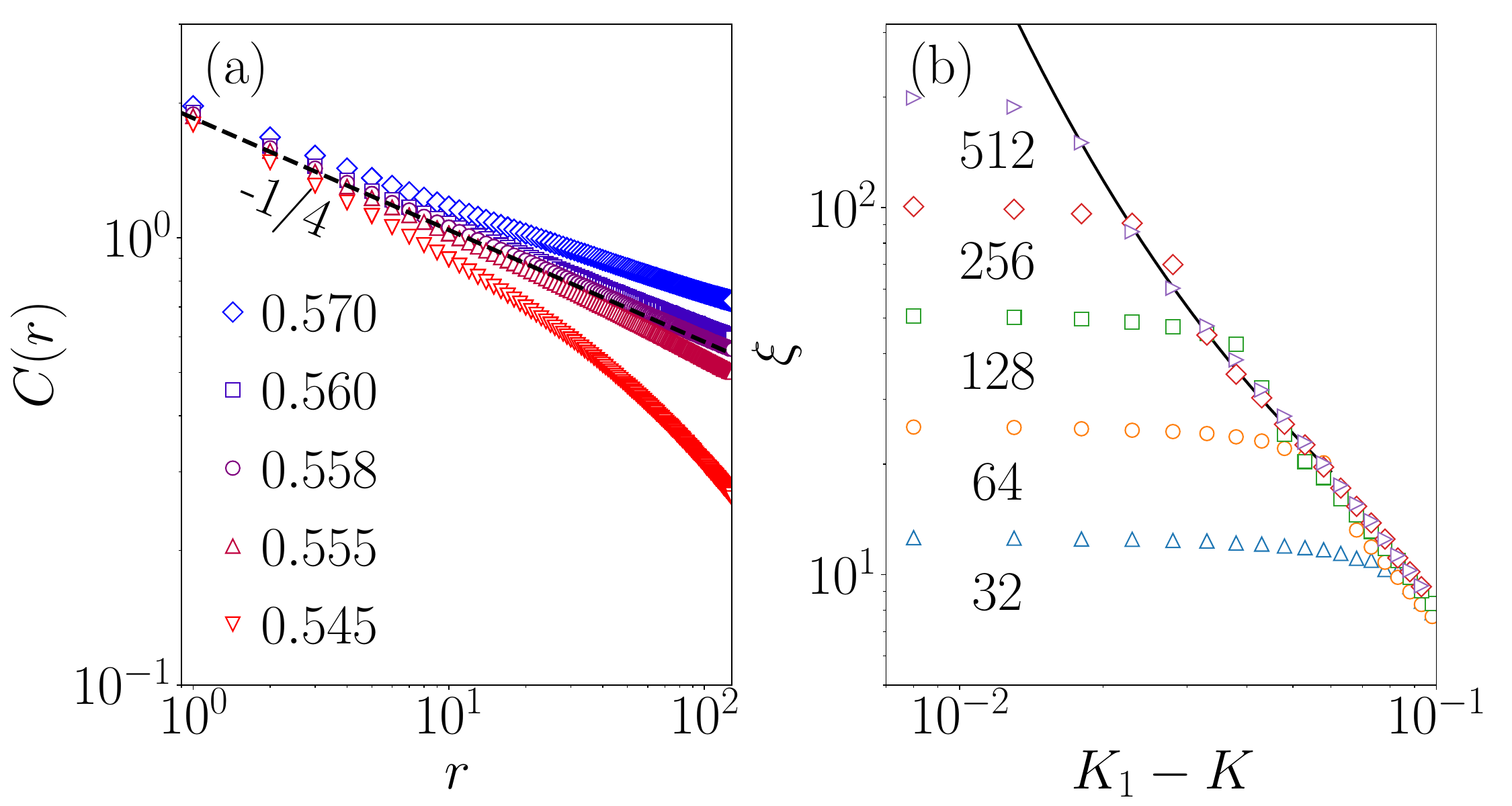}
    \caption{ (a) The spin-spin correlation function and (b) the correlation length near the disordered-QLRO phase transition point $K=K_1=0.558$ at $p=8$ and $\rho=2$. The correlation function follows a power law $C(r) \sim r^{-\eta}$ in the QLRO phase~($K \geq K_1$) with $\eta=1/4$ at the transition point.
    (b) The correlation length estimated using Eq.~\eqref{xi} grows as $K$ approaches the transition point and saturates to an $O(L)$ value when the correlation length exceeds the system size. The solid line is a curve $1.58 e^{0.611 |K-K_1|^{-1/2}}$ that fits the data at $L=512$ best. 
    }
    \label{fig7}
\end{figure}

The universality is further confirmed with the correlation function. We construct the cell spins $\bm{s}_\alpha = \sum_{i\in \alpha} \bm{u}_i$ with $b=1$ and  measure the correlation function $C(\bm{r})$
\begin{equation}
    C(\bm{r})= \left[ \left\langle \left( \bm{s}_\alpha - \bar{\bm{s}}\right)\cdot \left(\bm{s}_{\alpha'} - \bar{\bm{s}}\right) \right\rangle \right]_{\bm{r}_{\alpha'} - \bm{r}_\alpha = \bm{r}},
\end{equation}
where $\bar{\bm{s}} = \frac{1}{L^d}\sum_\alpha \bm{s}_\alpha$, the angular bracket denotes the steady state time average, and the square bracket denotes the average over all pairs of cell spins displaced by $\bm{r}$. Figure~\ref{fig7}(a) presents the correlation function near $K=K_1 \simeq 0.558$ at $\rho=2$. Inside the QLRO phase~($K>K_1$), the correlation function decays algebraically with the distance. The decay exponent at the transition point is in good agreement with the universal value $1/4$ of the vortex-antivortex unbinding BKT transition~\cite{Kosterlitz.1973}. 

Another hallmark of the BKT transition is the essential singularity~\cite{jose1977renormalization, Kosterlitz.1973, Izyumov.1988, Noh.2002} in the correlation length 
\begin{equation}\label{xi_essential}
    \xi(K) \sim e^{a|K-K_1|^{-1/2}}
\end{equation}
as the transition point is approached from the disordered phase.
The correlation length estimated as~\cite{Noh.2002}
\begin{equation}\label{xi}
    \xi = \sqrt{ \frac{\sum_{\bm{r}} |\bm{r}|^2 C(\bm{r})}{\sum_{\bm{r}} C(\bm{r})} } . 
\end{equation}
is presented in Fig.~\ref{fig7}(b). As $K$ increases, the correlation length increases until it saturates to a value $O(L)$. It is not decisive yet even at $L=512$ whether the growth of the correlation length follows the scaling form in Eq.~\eqref{xi_essential}. However, an upward curvature in the correlation length plot in the log-log scale suggests that the essential singularity would show up at larger systems. 

Our numerical results coherently suggest that the vortex-antivortex unbinding BKT transition is robust against the particle diffusion. We present a theoretical argument for the robustness.
In order to understand the role of the particle diffusion, we consider an
effective dynamics of a vortex and antivortex pair. Let $\bm{r}$ be the
relative displacement between the pair. 
{In the equilibrium system,} the pair interacts through the attracting logarithmic
potential $V(\bm{r}) \simeq 2\pi J \ln |\bm{r}|$ with a
{renormalized} coupling constant
$J$~\cite{Kosterlitz.1973,Izyumov.1988}. 
Thus, the time evolution is governed by the Langevin equation in dimensionless unit
\begin{equation}\label{vortex_dynamics}
    \dot{\bm{r}} = -\bm{\nabla} V(\bm{r}) + \sqrt{2T} \bm{\zeta}_{th}(t),
\end{equation}
where $\bm{\zeta}_{th}(t)$ denotes the $\delta$-correlated white noise of
unit variance {coming from stochastic thermal fluctuations of
spins.  The Langevin equation results in the power law distribution
$P_{eq}(\bm{r}) \sim e^{-\beta V(\bm{r})} \sim r^{-2\pi J/T}$ in
equilibrium~\cite{Kosterlitz.1973,Izyumov.1988,Goldenfeld.1992}. }
As a result of the competition between the logarithmic attraction and the
thermal fluctuation, the pair forms a bound state {with finite
$\langle r^2\rangle$} at low temperatures ($T<\pi J/2$) and becomes unbound at high temperatures. In the latter case, free vortices and antivortices proliferate and destroy the criticality. The binding-unbinding of the pair is the underlying mechanism for the BKT transition~\cite{Kosterlitz.1973}.

The particle diffusion in the Brownian clock model introduces an additional
diffusion noise. {Independent random walks of particles will
    induces additional diffusion motions of the vortex/antivortex cores.
    We note that spin fluctuations and diffusive fluctuations are completely
uncorrelated.} Thus, the effective Langevin equation for the vortex-antivortex pair would be given by 
\begin{equation}\label{Langevin_mod}
    \dot{\bm{r}} = -\bm{\nabla} V(\bm{r}) + \sqrt{2T} \bm{\zeta}_{th}(t) + \sqrt{2D} \bm{\zeta}_{d}(t),
\end{equation}
where $\bm{\zeta}_{d}(t)$ is $\delta$-correlated white noise of unit
variance representing the diffusion noise and $D\propto l_0^2$ is a
diffusion constant. The thermal noise and the diffusion noise are
uncorrelated for the passive Brownian particles. Thus, their sum acts as
single thermal noise with an effective temperature $T_{\rm eff} = T+D$.
In this argument, the particle diffusion only modifies the effective temperature of the topological excitation. Therefore, we conclude that the phase transition between the QLRO phase and the disordered phase belongs to the BKT universality class.

\begin{figure}[t]
    \includegraphics[width=1\linewidth]{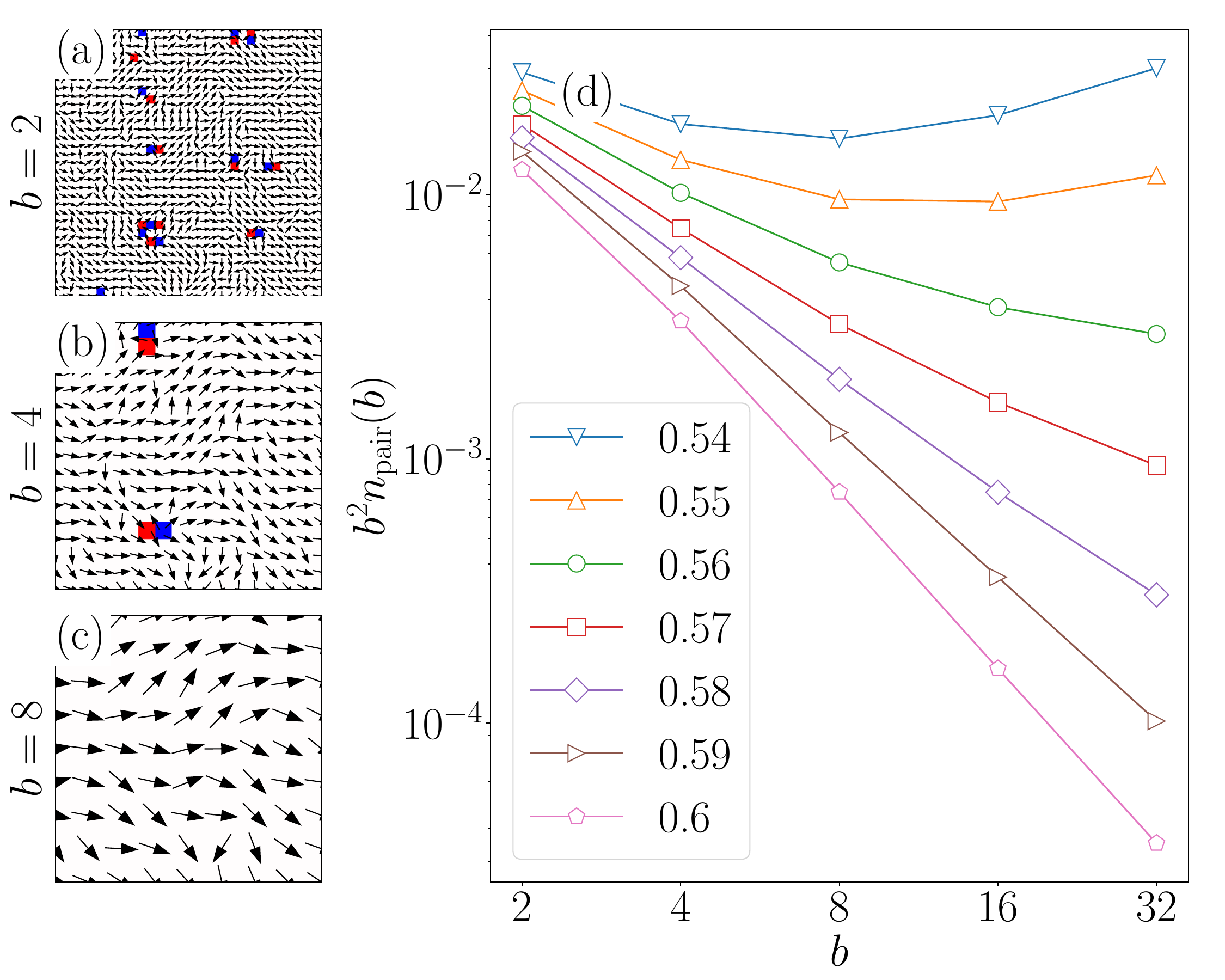}
    \caption{ The part of the spin configuration shown in Fig.~\ref{fig1}(c) are coarse-grained with cell sizes $b=2$ (a),
        $4$ (b), and $8$ (c). As $b$ increases, vortex-antivortex pairs at
        shorter distances are coarse-grained out. (d) The number density
        $n_{\rm pair}(b)$ of
        vortex-antivortex pairs multiplied with $b^2$ is plotted as a function of the
        coarse-graining cell size $b$ in the log-log scale. 
        The data $b^2 n_{\rm pair}(b)$ display the power law decay in the QLRO phase~($K<K_1\sim
        0.558$) while they increase in the disordered phase~($K>K_1$). These
        behaviors are consistent with the theoretical prediction explained in the main
        text.
        All the data were taken for the Brownian eight-state clock model with
        $\rho=2$ and $L=1024$.
    }
    \label{fig8}
\end{figure}

{ The Langevin equation in Eq.~\eqref{Langevin_mod} predicts that the probability distribution for the displacement
    is given by $P(\bm{r}) \sim e^{-V(\bm{r})/T_{\rm eff}} \sim r^{-\delta}$
    where $\delta = 2\pi J/T_{\rm
    eff}$~\cite{Kosterlitz.1973,Izyumov.1988,Goldenfeld.1992}. The exponent
    $\delta$ depends on the microscopic model parameters such as $K$ and
    $\rho$. 
    As a numerical evidence for the proposed Langevin equation, we measure the number density of the
    vortex-antivortex pairs as a function of the distance. 
    We coarse-grain a spin configuration using $b\times b$ cells and count
    the number of vortex-antivortex pair per unit area. 
    Under the coarse-graining, any vortex-antivortex pairs whose separation
    is smaller than $b$ are coarse-grained out and invisible~(see
    Fig.~\ref{fig8}(a,b,c)). Thus, the measured number density
    is given by
    \begin{equation}
        n_{\rm pair}(b) = \int_{|\bm{r}| \geq b} P(\bm{r}) d^2 \bm{r} .
        \label{n_pair}
    \end{equation}
Using $P(\bm{r})\sim r^{-\delta}$ with $\delta \geq 4$ in the QLRO phase, we
predict that $n_{\rm pair}(b) \sim r^{-(\delta-2)}$ decays algebraically
with $b$. In Fig.~\eqref{fig8}(c), we present the numerical data obtained
for the Brownian eight-state clock model with $\rho=2$ near the
vortex-antivortex unbinding transition point $K = K_1 \simeq 0.558$. The
number density follows the power law in the QLRO phase, which justifies the
Langevin equation proposed in Eq.~\eqref{Langevin_mod}.
}

\subsection{Energy flow as a nonequilibrium effect at the symmetry breaking transition}\label{Heat flow}

{As mentioned in Sec.~\ref{sec:model},
the Brownian clock model could be regarded as a system that is thermal
contact with the spin bath and the diffusion bath.}
In this section, we examine the energy flow between the two baths through the Brownian clock spin system near the symmetry breaking transition.

We introduce a local energy $e_i$ of particle $i$:
\begin{equation}
    e_i \equiv -\sum_{j \in \mathcal{N}_i} \cos(\theta_i - \theta_j),
\end{equation}
where $\mathcal{N}_i$ denotes the set of particles within the interaction range $r_0$ from $i$. Its value varies stochastically as spins flip and particles move. 
In order to separate the effects of both dynamics, we measure the probability distributions $P_{\rm post/pre}(e;n) = \langle \delta(e_i - e)\rangle_{{\rm post/pre}, n}$ of the local energy of a particle having $n$ interaction neighbors right before~($P_{\rm pre}$) and after~($P_{\rm post})$ the spin updates.

\begin{figure}[t]
    \includegraphics[width=1\linewidth]{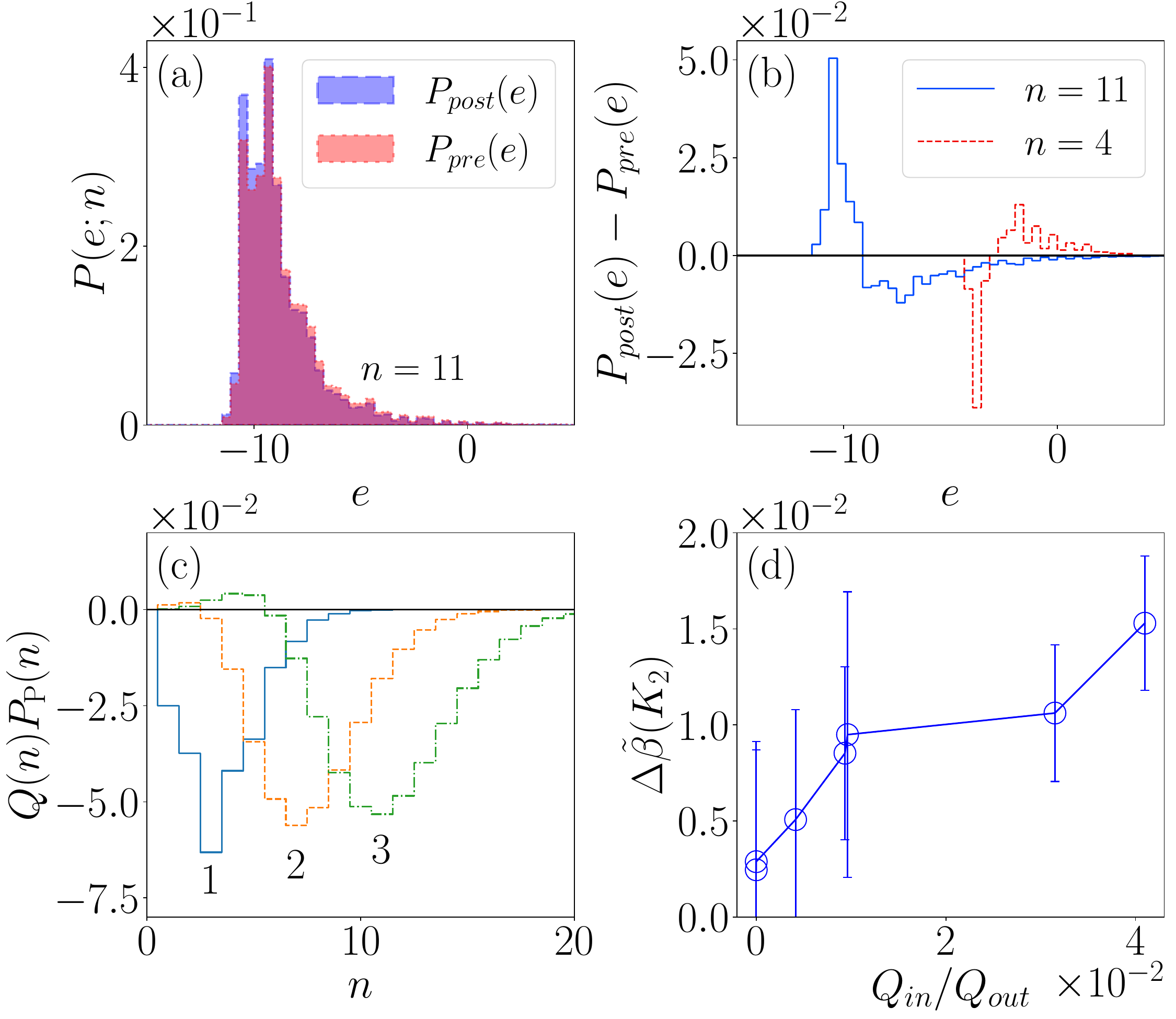}
    \caption{Energy flow in the Brownian eight-state clock model with $L=512$.
    (a) Local energy distributions $P_{\rm pre, post}(e;n=11)$ at $K=K_2=0.56$ when $\rho=3$. 
    (b) $P_{\rm pre}(e;n)-P_{\rm post}(e;n)$ when $n=11$ and $n=4$. Other parameters are the same as in (a). 
    (c) Average energy gain from the spin bath at the symmetry-breaking transition point $K=K_2$ at different values of the overall particle density $\rho=1, 2, 3$. 
    (d) Comparison of $\Delta \tilde\beta = 2 /p^2 - \tilde \beta (K_2)$ and the ratio $Q_{\rm in}/Q_{\rm out}$.
    }
    \label{fig9}
\end{figure}

Figures~\ref{fig9}(a) and (b) compare the probability distributions at the symmetry-breaking transition point $K=K_2$ when $\rho=3$. A particle in a dense region~($n=11$) loses an energy to the spin bath while one in a dilute region~($n=4$) gains an energy from the spin bath.
The average energy gain from the spin bath, or equivalently the energy loss to the diffusion bath, is given by
\begin{equation}\label{Eflow}
Q(n) = \int e P_{\rm post}(e;n) de - \int e P_{\rm pre}(e;n) de .
\end{equation}
The energy gain spectrum $Q(n)$ at the symmetry-breaking transition point $K=K_2$ is presented in Fig.~\ref{fig9}(c). We notice a qualitative difference in the energy gain spectrum depending on the overall particle density. 
When $\rho=3$, the system gains an energy from the spin bath~($Q(n)>0$) in dilute regions and loses an energy to the spin bath~($Q(n)<0$) in dense regions. The energy flow from the spin bath should be compensated by the counter energy flow from the diffusion bath. The sign change in $Q(n)$ indicates a spin energy transfer from the low density regions to the high density regions with the help of the diffusion bath. Such a spatial energy flow drives the system out of equilibrium. 
On the other hand, when $\rho=1$, particles gain an energy from the diffusion bath and lose it to the spin bath irrespective of the local density. Such an energy flow does not necessarily accompany a spatial energy flow.

As an indicator of the nonequilibrium driving, we propose to consider a
ratio $Q_{\rm in}/Q_{\rm out}$, where $Q_{\rm in}~(Q_{\rm out})$ is the
gross amount of the energy gain~(loss) of the system from~(to) the spin bath. They can be evaluated as
\begin{equation}
\begin{aligned}
    Q_{\rm in} & = \sum_{n} Q(n) \Theta(Q(n))P_{\rm P}(n;\rho) \\
    Q_{\rm out} & = \sum_{n} |Q(n)| \Theta(-Q(n)) P_{\rm P}(n;\rho),
\end{aligned}
\end{equation}
where $\Theta(x)$ is the Heaviside step function and $P_{P}(n;\rho) = e^{-n_0} \frac{n_0^n}{n!}$ is the Poisson distribution for the number of interacting neighbors with the mean $n_0 = \pi r_0^2 \rho$. Since particles diffuse freely, the number of particles within a circle of radius $r_0$ follows the Poisson distribution. 
In Fig.~\ref{fig9}(d), we compare the ratio $Q_{\rm in}/Q_{\rm out}$ and the deviation $\Delta \tilde{\beta} = 2/p^2 - \tilde\beta(K_2)$ of the finite-size-scaling exponent at the symmetry-breaking transition point from the equilibrium value. There is a positive correlation between them, which suggests that the nonequilibrium energy flow is responsible for the quantitative deviation of $\tilde{\beta}(K_2)$ from the equilibrium BKT transition picture. 

\begin{figure}[t]
    \includegraphics[width=1\linewidth]{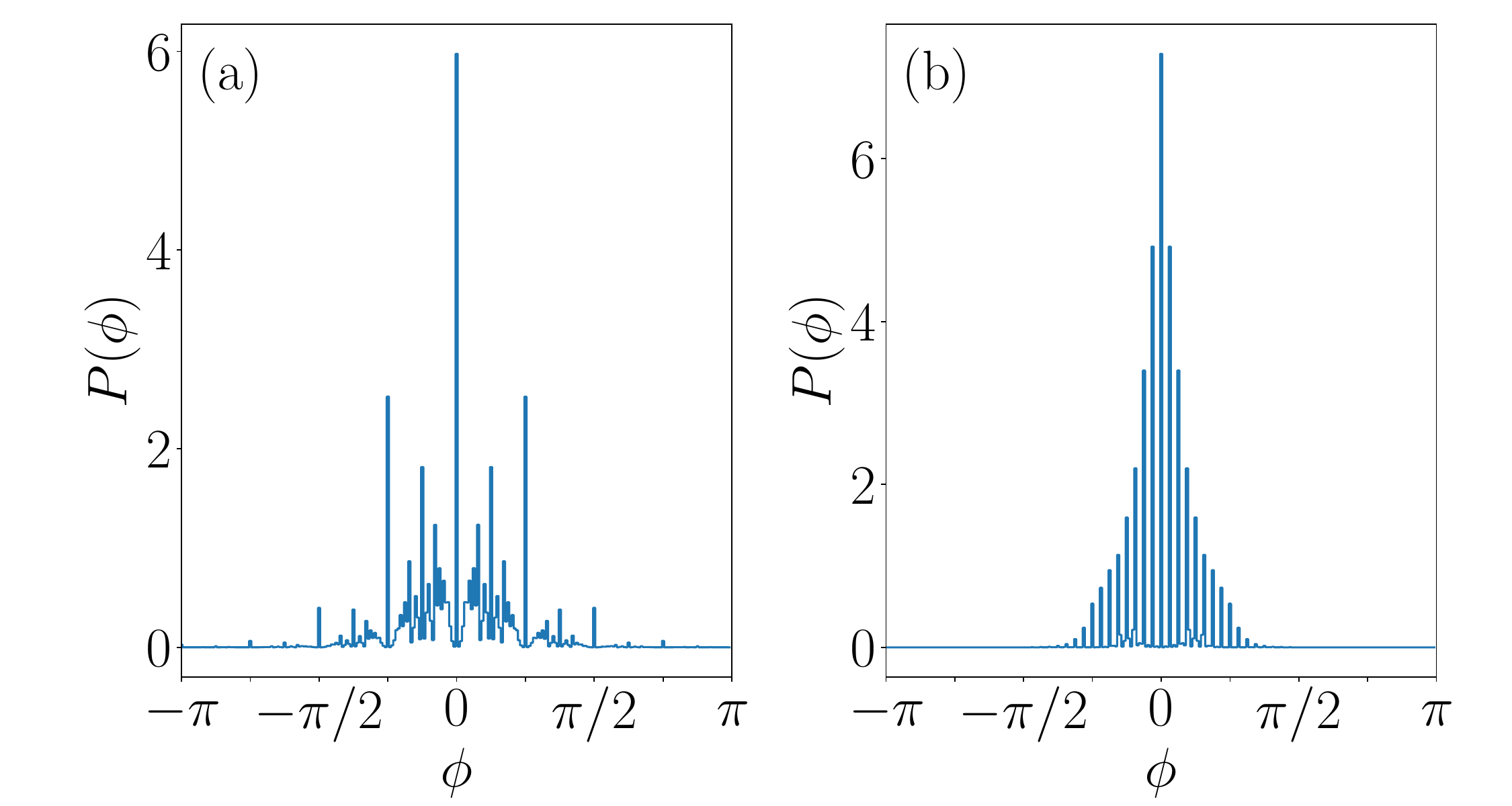}
    \caption{The probability distribution of the relative angle of a local
        field with respect to a central spin evaluated at the
        symmetry-breaking transition point of the Brownian clock model
        ($p=8, \rho=3, K=K_2 = 0.56, L=512$) in (a) and of the equilibrium
        lattice clock model~($p=8, J/k_BT = 2.4, L=512$) in (b). The average
        number of interaction neighbors is $n_0 = \pi r_0^2 \rho \simeq 9.4$
        in the Brownian clock model. The local field in the equilibrium
        model is calculated from eight spins from nearest and next nearest neighbors of a central spin. 
    }
    \label{fig10}
\end{figure}

The energy flow disturbs a local field $\bm{h}_i = \sum_{j\in \mathcal{N}_i}
{\bm u}_j$ around a spin ${\bm u}_i$ from the corresponding equilibrium
distribution. We characterize the nonequilibrium effect using the
distribution function $P(\phi) = \langle \delta(\phi_i - \phi)\rangle$ of
the relative angle $\phi_i \equiv {\rm arg}\left[\bm{h}_i \right] -
\theta_i$ of the local field.  The distribution function $P(\phi)$ evaluated
at the symmetry-breaking transition point $K=K_2$ at $\rho = 3$ is  compared
with the distribution function in the equilibrium lattice model in
Fig.~\ref{fig10}. The two distribution functions have significantly
different shapes, which signals the nonequilibrium effect in the Brownian
clock model. The former is characterized with intermittent peaks and is more widely distributed than the latter. 
Based on the complex distribution function, we further speculate that the
spin fluctuation in the Brownian $p$-state clock model might be described
with an effective value $p_{\rm eff.}$, which is larger than the bare value $p$. Were it true, the finite-size-scaling exponent at the symmetry-breaking transition would be given by $\tilde\beta = 2/p_{\rm eff.}^2 \leq 2/p^2$ as found in the numerical study. It will be interesting to develop a theory for $P(\phi)$ and $p_{\rm eff.}$, which is beyond the scope of the present study. 

\section{SUMMARY AND DISCUSSION} \label{sec:summary}

We have investigated the nature of the phase transitions in the Brownian $p$-state clock model with extensive numerical simulations. Our numerical results are summarized with the phase diagram in Fig.~\ref{fig2}. Despite the particle diffusion, the phase diagram has the same overall structure as that of the equilibrium model in the two-dimensional lattice.
When $p\leq 4$, the system undergoes a single order-disorder phase transition. When $4<p<\infty$, there appears the QLRO phase in between the ordered and disordered phases. The QLRO phase is a critical phase where the correlation function and the order parameter exhibit the power law scaling $C(r) \sim r^{-\eta}$ and $m \sim L^{-\tilde\beta}$ with the continuously varying critical exponents $\eta$ and $\tilde\beta$ satisfying the scaling relation $\tilde\beta = \eta/2$.

We found that the vortex-antivortex unbinding BKT transition is robust against the particle diffusion. Our numerical results show that $\tilde\beta$ takes the universal value $1/8$ at the transition to the disordered phase. We presented a theoretical argument based on the effective dynamics of vortex-antivortex pairs. The particle diffusion only modifies the effective temperature for the topological excitations, hence it only shifts the transition point and does not alter the transition nature. 

On the other hand, it is questionable whether the symmetry-breaking phase transition from the QLRO phase to the ordered phase also belongs to the equilibrium BKT universality class. The equilibrium theory predicts that $\tilde\beta$ takes the universal value $2/p^2$ at the transition point. On the contrary, $\tilde\beta(K_2)$ in the Brownian clock model is nonuniversal: It deviates from $2/p^2$ and the discrepancy increases systematically as the particle density increases~(see Fig.~\ref{fig6} (b)). We observed that the particle diffusion induces an energy flow through the system, which drives the system out of thermal equilibrium. Quantitatively, the energy flow is positively correlated with the discrepancy $\Delta \tilde\beta(K_2) = 2/p^2 - \tilde\beta(K_2)$~(see Fig.~\ref{fig9}(d)). Based on this observation, we conjecture that the nonequilibrium heat flow is responsible for the nonuniversal BKT transition to the ordered phase.
The discrepancy $\tilde\beta(K_2) < 2/p^2$ might indicate a change in the number of spin states from the bare value $p$ to a renormalized value $p_{\rm eff.} \geq p$.
It would be interesting to examine such a possibility, which we leave for a future work.

\begin{acknowledgments}
  This work is supported by the 2021 Research Fund of the University of Seoul.
  We also acknowledge the computing resources of Urban Big data and AI Institute (UBAI) at the University of Seoul.
\end{acknowledgments}

\bibliography{paper}

\begin{thebibliography}{37}%
\makeatletter
\providecommand \@ifxundefined [1]{%
 \@ifx{#1\undefined}
}%
\providecommand \@ifnum [1]{%
 \ifnum #1\expandafter \@firstoftwo
 \else \expandafter \@secondoftwo
 \fi
}%
\providecommand \@ifx [1]{%
 \ifx #1\expandafter \@firstoftwo
 \else \expandafter \@secondoftwo
 \fi
}%
\providecommand \natexlab [1]{#1}%
\providecommand \enquote  [1]{``#1''}%
\providecommand \bibnamefont  [1]{#1}%
\providecommand \bibfnamefont [1]{#1}%
\providecommand \citenamefont [1]{#1}%
\providecommand \href@noop [0]{\@secondoftwo}%
\providecommand \href [0]{\begingroup \@sanitize@url \@href}%
\providecommand \@href[1]{\@@startlink{#1}\@@href}%
\providecommand \@@href[1]{\endgroup#1\@@endlink}%
\providecommand \@sanitize@url [0]{\catcode `\\12\catcode `\$12\catcode
  `\&12\catcode `\#12\catcode `\^12\catcode `\_12\catcode `\%12\relax}%
\providecommand \@@startlink[1]{}%
\providecommand \@@endlink[0]{}%
\providecommand \url  [0]{\begingroup\@sanitize@url \@url }%
\providecommand \@url [1]{\endgroup\@href {#1}{\urlprefix }}%
\providecommand \urlprefix  [0]{URL }%
\providecommand \Eprint [0]{\href }%
\providecommand \doibase [0]{https://doi.org/}%
\providecommand \selectlanguage [0]{\@gobble}%
\providecommand \bibinfo  [0]{\@secondoftwo}%
\providecommand \bibfield  [0]{\@secondoftwo}%
\providecommand \translation [1]{[#1]}%
\providecommand \BibitemOpen [0]{}%
\providecommand \bibitemStop [0]{}%
\providecommand \bibitemNoStop [0]{.\EOS\space}%
\providecommand \EOS [0]{\spacefactor3000\relax}%
\providecommand \BibitemShut  [1]{\csname bibitem#1\endcsname}%
\let\auto@bib@innerbib\@empty
\bibitem [{\citenamefont {Marchetti}\ \emph {et~al.}(2013)\citenamefont
  {Marchetti}, \citenamefont {Joanny}, \citenamefont {Ramaswamy}, \citenamefont
  {Liverpool}, \citenamefont {Prost}, \citenamefont {Rao},\ and\ \citenamefont
  {Simha}}]{marchetti2013hydrodynamics}%
  \BibitemOpen
  \bibfield  {author} {\bibinfo {author} {\bibfnamefont {M.~C.}\ \bibnamefont
  {Marchetti}}, \bibinfo {author} {\bibfnamefont {J.-F.}\ \bibnamefont
  {Joanny}}, \bibinfo {author} {\bibfnamefont {S.}~\bibnamefont {Ramaswamy}},
  \bibinfo {author} {\bibfnamefont {T.~B.}\ \bibnamefont {Liverpool}}, \bibinfo
  {author} {\bibfnamefont {J.}~\bibnamefont {Prost}}, \bibinfo {author}
  {\bibfnamefont {M.}~\bibnamefont {Rao}},\ and\ \bibinfo {author}
  {\bibfnamefont {R.~A.}\ \bibnamefont {Simha}},\ }\bibfield  {title} {\bibinfo
  {title} {Hydrodynamics of soft active matter},\ }\href@noop {} {\bibfield
  {journal} {\bibinfo  {journal} {Reviews of modern physics}\ }\textbf
  {\bibinfo {volume} {85}},\ \bibinfo {pages} {1143} (\bibinfo {year}
  {2013})}\BibitemShut {NoStop}%
\bibitem [{\citenamefont {Elgeti}\ \emph {et~al.}(2015)\citenamefont {Elgeti},
  \citenamefont {Winkler},\ and\ \citenamefont {Gompper}}]{elgeti2015physics}%
  \BibitemOpen
  \bibfield  {author} {\bibinfo {author} {\bibfnamefont {J.}~\bibnamefont
  {Elgeti}}, \bibinfo {author} {\bibfnamefont {R.~G.}\ \bibnamefont
  {Winkler}},\ and\ \bibinfo {author} {\bibfnamefont {G.}~\bibnamefont
  {Gompper}},\ }\bibfield  {title} {\bibinfo {title} {Physics of
  microswimmers—single particle motion and collective behavior: a review},\
  }\href@noop {} {\bibfield  {journal} {\bibinfo  {journal} {Reports on
  progress in physics}\ }\textbf {\bibinfo {volume} {78}},\ \bibinfo {pages}
  {056601} (\bibinfo {year} {2015})}\BibitemShut {NoStop}%
\bibitem [{\citenamefont {Bechinger}\ \emph {et~al.}(2016)\citenamefont
  {Bechinger}, \citenamefont {Di~Leonardo}, \citenamefont {L{\"o}wen},
  \citenamefont {Reichhardt}, \citenamefont {Volpe},\ and\ \citenamefont
  {Volpe}}]{bechinger2016active}%
  \BibitemOpen
  \bibfield  {author} {\bibinfo {author} {\bibfnamefont {C.}~\bibnamefont
  {Bechinger}}, \bibinfo {author} {\bibfnamefont {R.}~\bibnamefont
  {Di~Leonardo}}, \bibinfo {author} {\bibfnamefont {H.}~\bibnamefont
  {L{\"o}wen}}, \bibinfo {author} {\bibfnamefont {C.}~\bibnamefont
  {Reichhardt}}, \bibinfo {author} {\bibfnamefont {G.}~\bibnamefont {Volpe}},\
  and\ \bibinfo {author} {\bibfnamefont {G.}~\bibnamefont {Volpe}},\ }\bibfield
   {title} {\bibinfo {title} {Active particles in complex and crowded
  environments},\ }\href@noop {} {\bibfield  {journal} {\bibinfo  {journal}
  {Reviews of Modern Physics}\ }\textbf {\bibinfo {volume} {88}},\ \bibinfo
  {pages} {045006} (\bibinfo {year} {2016})}\BibitemShut {NoStop}%
\bibitem [{\citenamefont {Cates}\ and\ \citenamefont
  {Tailleur}(2015)}]{cates2015motility}%
  \BibitemOpen
  \bibfield  {author} {\bibinfo {author} {\bibfnamefont {M.~E.}\ \bibnamefont
  {Cates}}\ and\ \bibinfo {author} {\bibfnamefont {J.}~\bibnamefont
  {Tailleur}},\ }\bibfield  {title} {\bibinfo {title} {Motility-induced phase
  separation},\ }\href@noop {} {\bibfield  {journal} {\bibinfo  {journal}
  {Annu. Rev. Condens. Matter Phys.}\ }\textbf {\bibinfo {volume} {6}},\
  \bibinfo {pages} {219} (\bibinfo {year} {2015})}\BibitemShut {NoStop}%
\bibitem [{\citenamefont {Di~Leonardo}\ \emph {et~al.}(2010)\citenamefont
  {Di~Leonardo}, \citenamefont {Angelani}, \citenamefont {Dell’Arciprete},
  \citenamefont {Ruocco}, \citenamefont {Iebba}, \citenamefont {Schippa},
  \citenamefont {Conte}, \citenamefont {Mecarini}, \citenamefont {De~Angelis},\
  and\ \citenamefont {Di~Fabrizio}}]{di2010bacterial}%
  \BibitemOpen
  \bibfield  {author} {\bibinfo {author} {\bibfnamefont {R.}~\bibnamefont
  {Di~Leonardo}}, \bibinfo {author} {\bibfnamefont {L.}~\bibnamefont
  {Angelani}}, \bibinfo {author} {\bibfnamefont {D.}~\bibnamefont
  {Dell’Arciprete}}, \bibinfo {author} {\bibfnamefont {G.}~\bibnamefont
  {Ruocco}}, \bibinfo {author} {\bibfnamefont {V.}~\bibnamefont {Iebba}},
  \bibinfo {author} {\bibfnamefont {S.}~\bibnamefont {Schippa}}, \bibinfo
  {author} {\bibfnamefont {M.~P.}\ \bibnamefont {Conte}}, \bibinfo {author}
  {\bibfnamefont {F.}~\bibnamefont {Mecarini}}, \bibinfo {author}
  {\bibfnamefont {F.}~\bibnamefont {De~Angelis}},\ and\ \bibinfo {author}
  {\bibfnamefont {E.}~\bibnamefont {Di~Fabrizio}},\ }\bibfield  {title}
  {\bibinfo {title} {Bacterial ratchet motors},\ }\href@noop {} {\bibfield
  {journal} {\bibinfo  {journal} {Proceedings of the National Academy of
  Sciences}\ }\textbf {\bibinfo {volume} {107}},\ \bibinfo {pages} {9541}
  (\bibinfo {year} {2010})}\BibitemShut {NoStop}%
\bibitem [{\citenamefont {Cavagna}\ and\ \citenamefont
  {Giardina}(2014)}]{cavagna2014bird}%
  \BibitemOpen
  \bibfield  {author} {\bibinfo {author} {\bibfnamefont {A.}~\bibnamefont
  {Cavagna}}\ and\ \bibinfo {author} {\bibfnamefont {I.}~\bibnamefont
  {Giardina}},\ }\bibfield  {title} {\bibinfo {title} {Bird flocks as condensed
  matter},\ }\href@noop {} {\bibfield  {journal} {\bibinfo  {journal} {Annu.
  Rev. Condens. Matter Phys.}\ }\textbf {\bibinfo {volume} {5}},\ \bibinfo
  {pages} {183} (\bibinfo {year} {2014})}\BibitemShut {NoStop}%
\bibitem [{\citenamefont {Ballerini}\ \emph {et~al.}(2008)\citenamefont
  {Ballerini}, \citenamefont {Cabibbo}, \citenamefont {Candelier},
  \citenamefont {Cavagna}, \citenamefont {Cisbani}, \citenamefont {Giardina},
  \citenamefont {Lecomte}, \citenamefont {Orlandi}, \citenamefont {Parisi},
  \citenamefont {Procaccini} \emph {et~al.}}]{ballerini2008interaction}%
  \BibitemOpen
  \bibfield  {author} {\bibinfo {author} {\bibfnamefont {M.}~\bibnamefont
  {Ballerini}}, \bibinfo {author} {\bibfnamefont {N.}~\bibnamefont {Cabibbo}},
  \bibinfo {author} {\bibfnamefont {R.}~\bibnamefont {Candelier}}, \bibinfo
  {author} {\bibfnamefont {A.}~\bibnamefont {Cavagna}}, \bibinfo {author}
  {\bibfnamefont {E.}~\bibnamefont {Cisbani}}, \bibinfo {author} {\bibfnamefont
  {I.}~\bibnamefont {Giardina}}, \bibinfo {author} {\bibfnamefont
  {V.}~\bibnamefont {Lecomte}}, \bibinfo {author} {\bibfnamefont
  {A.}~\bibnamefont {Orlandi}}, \bibinfo {author} {\bibfnamefont
  {G.}~\bibnamefont {Parisi}}, \bibinfo {author} {\bibfnamefont
  {A.}~\bibnamefont {Procaccini}}, \emph {et~al.},\ }\bibfield  {title}
  {\bibinfo {title} {Interaction ruling animal collective behavior depends on
  topological rather than metric distance: Evidence from a field study},\
  }\href@noop {} {\bibfield  {journal} {\bibinfo  {journal} {Proceedings of the
  national academy of sciences}\ }\textbf {\bibinfo {volume} {105}},\ \bibinfo
  {pages} {1232} (\bibinfo {year} {2008})}\BibitemShut {NoStop}%
\bibitem [{\citenamefont {Ward}\ \emph {et~al.}(2008)\citenamefont {Ward},
  \citenamefont {Sumpter}, \citenamefont {Couzin}, \citenamefont {Hart},\ and\
  \citenamefont {Krause}}]{ward2008quorum}%
  \BibitemOpen
  \bibfield  {author} {\bibinfo {author} {\bibfnamefont {A.~J.}\ \bibnamefont
  {Ward}}, \bibinfo {author} {\bibfnamefont {D.~J.}\ \bibnamefont {Sumpter}},
  \bibinfo {author} {\bibfnamefont {I.~D.}\ \bibnamefont {Couzin}}, \bibinfo
  {author} {\bibfnamefont {P.~J.}\ \bibnamefont {Hart}},\ and\ \bibinfo
  {author} {\bibfnamefont {J.}~\bibnamefont {Krause}},\ }\bibfield  {title}
  {\bibinfo {title} {Quorum decision-making facilitates information transfer in
  fish shoals},\ }\href@noop {} {\bibfield  {journal} {\bibinfo  {journal}
  {Proceedings of the National Academy of Sciences}\ }\textbf {\bibinfo
  {volume} {105}},\ \bibinfo {pages} {6948} (\bibinfo {year}
  {2008})}\BibitemShut {NoStop}%
\bibitem [{\citenamefont {Cavagna}\ \emph {et~al.}(2017)\citenamefont
  {Cavagna}, \citenamefont {Conti}, \citenamefont {Creato}, \citenamefont
  {Del~Castello}, \citenamefont {Giardina}, \citenamefont {Grigera},
  \citenamefont {Melillo}, \citenamefont {Parisi},\ and\ \citenamefont
  {Viale}}]{cavagna2017dynamic}%
  \BibitemOpen
  \bibfield  {author} {\bibinfo {author} {\bibfnamefont {A.}~\bibnamefont
  {Cavagna}}, \bibinfo {author} {\bibfnamefont {D.}~\bibnamefont {Conti}},
  \bibinfo {author} {\bibfnamefont {C.}~\bibnamefont {Creato}}, \bibinfo
  {author} {\bibfnamefont {L.}~\bibnamefont {Del~Castello}}, \bibinfo {author}
  {\bibfnamefont {I.}~\bibnamefont {Giardina}}, \bibinfo {author}
  {\bibfnamefont {T.~S.}\ \bibnamefont {Grigera}}, \bibinfo {author}
  {\bibfnamefont {S.}~\bibnamefont {Melillo}}, \bibinfo {author} {\bibfnamefont
  {L.}~\bibnamefont {Parisi}},\ and\ \bibinfo {author} {\bibfnamefont
  {M.}~\bibnamefont {Viale}},\ }\bibfield  {title} {\bibinfo {title} {Dynamic
  scaling in natural swarms},\ }\href@noop {} {\bibfield  {journal} {\bibinfo
  {journal} {Nature Physics}\ }\textbf {\bibinfo {volume} {13}},\ \bibinfo
  {pages} {914} (\bibinfo {year} {2017})}\BibitemShut {NoStop}%
\bibitem [{\citenamefont {Kumar}\ \emph {et~al.}(2014)\citenamefont {Kumar},
  \citenamefont {Soni}, \citenamefont {Ramaswamy},\ and\ \citenamefont
  {Sood}}]{kumar2014flocking}%
  \BibitemOpen
  \bibfield  {author} {\bibinfo {author} {\bibfnamefont {N.}~\bibnamefont
  {Kumar}}, \bibinfo {author} {\bibfnamefont {H.}~\bibnamefont {Soni}},
  \bibinfo {author} {\bibfnamefont {S.}~\bibnamefont {Ramaswamy}},\ and\
  \bibinfo {author} {\bibfnamefont {A.}~\bibnamefont {Sood}},\ }\bibfield
  {title} {\bibinfo {title} {Flocking at a distance in active granular
  matter},\ }\href@noop {} {\bibfield  {journal} {\bibinfo  {journal} {Nature
  communications}\ }\textbf {\bibinfo {volume} {5}},\ \bibinfo {pages} {4688}
  (\bibinfo {year} {2014})}\BibitemShut {NoStop}%
\bibitem [{\citenamefont {Vicsek}\ \emph {et~al.}(1995)\citenamefont {Vicsek},
  \citenamefont {Czir{\'o}k}, \citenamefont {Ben-Jacob}, \citenamefont
  {Cohen},\ and\ \citenamefont {Shochet}}]{vicsek1995novel}%
  \BibitemOpen
  \bibfield  {author} {\bibinfo {author} {\bibfnamefont {T.}~\bibnamefont
  {Vicsek}}, \bibinfo {author} {\bibfnamefont {A.}~\bibnamefont {Czir{\'o}k}},
  \bibinfo {author} {\bibfnamefont {E.}~\bibnamefont {Ben-Jacob}}, \bibinfo
  {author} {\bibfnamefont {I.}~\bibnamefont {Cohen}},\ and\ \bibinfo {author}
  {\bibfnamefont {O.}~\bibnamefont {Shochet}},\ }\bibfield  {title} {\bibinfo
  {title} {Novel type of phase transition in a system of self-driven
  particles},\ }\href@noop {} {\bibfield  {journal} {\bibinfo  {journal}
  {Physical Review Letters}\ }\textbf {\bibinfo {volume} {75}},\ \bibinfo
  {pages} {1226} (\bibinfo {year} {1995})}\BibitemShut {NoStop}%
\bibitem [{\citenamefont {Mermin}\ and\ \citenamefont
  {Wagner}(1966)}]{mermin1966absence}%
  \BibitemOpen
  \bibfield  {author} {\bibinfo {author} {\bibfnamefont {N.~D.}\ \bibnamefont
  {Mermin}}\ and\ \bibinfo {author} {\bibfnamefont {H.}~\bibnamefont
  {Wagner}},\ }\bibfield  {title} {\bibinfo {title} {Absence of ferromagnetism
  or antiferromagnetism in one-or two-dimensional isotropic heisenberg
  models},\ }\href@noop {} {\bibfield  {journal} {\bibinfo  {journal} {Physical
  Review Letters}\ }\textbf {\bibinfo {volume} {17}},\ \bibinfo {pages} {1133}
  (\bibinfo {year} {1966})}\BibitemShut {NoStop}%
\bibitem [{\citenamefont {Woo}\ \emph {et~al.}(2022)\citenamefont {Woo},
  \citenamefont {Rieger},\ and\ \citenamefont {Noh}}]{woo2022suppression}%
  \BibitemOpen
  \bibfield  {author} {\bibinfo {author} {\bibfnamefont {C.-U.}\ \bibnamefont
  {Woo}}, \bibinfo {author} {\bibfnamefont {H.}~\bibnamefont {Rieger}},\ and\
  \bibinfo {author} {\bibfnamefont {J.~D.}\ \bibnamefont {Noh}},\ }\bibfield
  {title} {\bibinfo {title} {Suppression of discontinuous phase transitions by
  particle diffusion},\ }\href@noop {} {\bibfield  {journal} {\bibinfo
  {journal} {Physical Review E}\ }\textbf {\bibinfo {volume} {105}},\ \bibinfo
  {pages} {054144} (\bibinfo {year} {2022})}\BibitemShut {NoStop}%
\bibitem [{\citenamefont {Ortiz}\ \emph {et~al.}(2012)\citenamefont {Ortiz},
  \citenamefont {Cobanera},\ and\ \citenamefont {Nussinov}}]{Ortiz.2012}%
  \BibitemOpen
  \bibfield  {author} {\bibinfo {author} {\bibfnamefont {G.}~\bibnamefont
  {Ortiz}}, \bibinfo {author} {\bibfnamefont {E.}~\bibnamefont {Cobanera}},\
  and\ \bibinfo {author} {\bibfnamefont {Z.}~\bibnamefont {Nussinov}},\
  }\bibfield  {title} {\bibinfo {title} {{Dualities and the phase diagram of
  the p-clock model}},\ }\href
  {https://doi.org/10.1016/j.nuclphysb.2011.09.012} {\bibfield  {journal}
  {\bibinfo  {journal} {Nuclear Physics B}\ }\textbf {\bibinfo {volume}
  {854}},\ \bibinfo {pages} {780} (\bibinfo {year} {2012})}\BibitemShut
  {NoStop}%
\bibitem [{\citenamefont {Li}\ \emph {et~al.}(2020)\citenamefont {Li},
  \citenamefont {Yang}, \citenamefont {Xie}, \citenamefont {Tu}, \citenamefont
  {Liao},\ and\ \citenamefont {Xiang}}]{Li.2020}%
  \BibitemOpen
  \bibfield  {author} {\bibinfo {author} {\bibfnamefont {Z.-Q.}\ \bibnamefont
  {Li}}, \bibinfo {author} {\bibfnamefont {L.-P.}\ \bibnamefont {Yang}},
  \bibinfo {author} {\bibfnamefont {Z.~Y.}\ \bibnamefont {Xie}}, \bibinfo
  {author} {\bibfnamefont {H.-H.}\ \bibnamefont {Tu}}, \bibinfo {author}
  {\bibfnamefont {H.-J.}\ \bibnamefont {Liao}},\ and\ \bibinfo {author}
  {\bibfnamefont {T.}~\bibnamefont {Xiang}},\ }\bibfield  {title} {\bibinfo
  {title} {{Critical properties of the two-dimensional q-state clock model}},\
  }\href {https://doi.org/10.1103/physreve.101.060105} {\bibfield  {journal}
  {\bibinfo  {journal} {Physical Review E}\ }\textbf {\bibinfo {volume}
  {101}},\ \bibinfo {pages} {060105} (\bibinfo {year} {2020})}\BibitemShut
  {NoStop}%
\bibitem [{\citenamefont {Kosterlitz}\ and\ \citenamefont
  {Thouless}(1973)}]{Kosterlitz.1973}%
  \BibitemOpen
  \bibfield  {author} {\bibinfo {author} {\bibfnamefont {J.~M.}\ \bibnamefont
  {Kosterlitz}}\ and\ \bibinfo {author} {\bibfnamefont {D.~J.}\ \bibnamefont
  {Thouless}},\ }\bibfield  {title} {\bibinfo {title} {{Ordering, metastability
  and phase transitions in two-dimensional systems}},\ }\href
  {https://doi.org/10.1088/0022-3719/6/7/010} {\bibfield  {journal} {\bibinfo
  {journal} {Journal of Physics C: Solid State Physics}\ }\textbf {\bibinfo
  {volume} {6}},\ \bibinfo {pages} {1181} (\bibinfo {year} {1973})}\BibitemShut
  {NoStop}%
\bibitem [{\citenamefont {Jos{\'e}}\ \emph {et~al.}(1977)\citenamefont
  {Jos{\'e}}, \citenamefont {Kadanoff}, \citenamefont {Kirkpatrick},\ and\
  \citenamefont {Nelson}}]{jose1977renormalization}%
  \BibitemOpen
  \bibfield  {author} {\bibinfo {author} {\bibfnamefont {J.~V.}\ \bibnamefont
  {Jos{\'e}}}, \bibinfo {author} {\bibfnamefont {L.~P.}\ \bibnamefont
  {Kadanoff}}, \bibinfo {author} {\bibfnamefont {S.}~\bibnamefont
  {Kirkpatrick}},\ and\ \bibinfo {author} {\bibfnamefont {D.~R.}\ \bibnamefont
  {Nelson}},\ }\bibfield  {title} {\bibinfo {title} {Renormalization, vortices,
  and symmetry-breaking perturbations in the two-dimensional planar model},\
  }\href@noop {} {\bibfield  {journal} {\bibinfo  {journal} {Physical Review
  B}\ }\textbf {\bibinfo {volume} {16}},\ \bibinfo {pages} {1217} (\bibinfo
  {year} {1977})}\BibitemShut {NoStop}%
\bibitem [{\citenamefont {Berezinskii}(1971)}]{berezinskii.1971}%
  \BibitemOpen
  \bibfield  {author} {\bibinfo {author} {\bibfnamefont {V.}~\bibnamefont
  {Berezinskii}},\ }\bibfield  {title} {\bibinfo {title} {Destruction of
  long-range order in one-dimensional and two-dimensional systems having a
  continuous symmetry group i. classical systems},\ }\href@noop {} {\bibfield
  {journal} {\bibinfo  {journal} {Sov. Phys. JETP}\ }\textbf {\bibinfo {volume}
  {32}},\ \bibinfo {pages} {493} (\bibinfo {year} {1971})}\BibitemShut
  {NoStop}%
\bibitem [{\citenamefont {Izyumov}\ and\ \citenamefont
  {Skryabin}(1988)}]{Izyumov.1988}%
  \BibitemOpen
  \bibfield  {author} {\bibinfo {author} {\bibfnamefont {Y.~A.}\ \bibnamefont
  {Izyumov}}\ and\ \bibinfo {author} {\bibfnamefont {Y.~N.}\ \bibnamefont
  {Skryabin}},\ }\href@noop {} {\emph {\bibinfo {title} {{Statistical Mechanics
  of Magnetically Ordered Systems}}}}\ (\bibinfo  {publisher}
  {Springer-Verlag},\ \bibinfo {address} {Berlin},\ \bibinfo {year}
  {1988})\BibitemShut {NoStop}%
\bibitem [{\citenamefont {Noh}\ \emph {et~al.}(2002)\citenamefont {Noh},
  \citenamefont {Rieger}, \citenamefont {Enderle},\ and\ \citenamefont
  {Knorr}}]{Noh.2002}%
  \BibitemOpen
  \bibfield  {author} {\bibinfo {author} {\bibfnamefont {J.~D.}\ \bibnamefont
  {Noh}}, \bibinfo {author} {\bibfnamefont {H.}~\bibnamefont {Rieger}},
  \bibinfo {author} {\bibfnamefont {M.}~\bibnamefont {Enderle}},\ and\ \bibinfo
  {author} {\bibfnamefont {K.}~\bibnamefont {Knorr}},\ }\bibfield  {title}
  {\bibinfo {title} {{Critical behavior of the frustrated antiferromagnetic
  six-state clock model on a triangular lattice}},\ }\href
  {https://doi.org/10.1103/physreve.66.026111} {\bibfield  {journal} {\bibinfo
  {journal} {Physical Review E}\ }\textbf {\bibinfo {volume} {66}},\ \bibinfo
  {pages} {026111} (\bibinfo {year} {2002})}\BibitemShut {NoStop}%
\bibitem [{\citenamefont {Tobochnik}(1982)}]{tobochnik1982properties}%
  \BibitemOpen
  \bibfield  {author} {\bibinfo {author} {\bibfnamefont {J.}~\bibnamefont
  {Tobochnik}},\ }\bibfield  {title} {\bibinfo {title} {Properties of the
  q-state clock model for q= 4, 5, and 6},\ }\href@noop {} {\bibfield
  {journal} {\bibinfo  {journal} {Physical Review B}\ }\textbf {\bibinfo
  {volume} {26}},\ \bibinfo {pages} {6201} (\bibinfo {year}
  {1982})}\BibitemShut {NoStop}%
\bibitem [{\citenamefont {Lapilli}\ \emph {et~al.}(2006)\citenamefont
  {Lapilli}, \citenamefont {Pfeifer},\ and\ \citenamefont
  {Wexler}}]{lapilli2006universality}%
  \BibitemOpen
  \bibfield  {author} {\bibinfo {author} {\bibfnamefont {C.~M.}\ \bibnamefont
  {Lapilli}}, \bibinfo {author} {\bibfnamefont {P.}~\bibnamefont {Pfeifer}},\
  and\ \bibinfo {author} {\bibfnamefont {C.}~\bibnamefont {Wexler}},\
  }\bibfield  {title} {\bibinfo {title} {Universality away from critical points
  in two-dimensional phase transitions},\ }\href@noop {} {\bibfield  {journal}
  {\bibinfo  {journal} {Physical Review Letters}\ }\textbf {\bibinfo {volume}
  {96}},\ \bibinfo {pages} {140603} (\bibinfo {year} {2006})}\BibitemShut
  {NoStop}%
\bibitem [{\citenamefont {Tuan}\ \emph {et~al.}(2022)\citenamefont {Tuan},
  \citenamefont {Long}, \citenamefont {Nui}, \citenamefont {Minh},
  \citenamefont {Kien},\ and\ \citenamefont {Viet}}]{tuan2022binder}%
  \BibitemOpen
  \bibfield  {author} {\bibinfo {author} {\bibfnamefont {L.~M.}\ \bibnamefont
  {Tuan}}, \bibinfo {author} {\bibfnamefont {T.~T.}\ \bibnamefont {Long}},
  \bibinfo {author} {\bibfnamefont {D.~X.}\ \bibnamefont {Nui}}, \bibinfo
  {author} {\bibfnamefont {P.~T.}\ \bibnamefont {Minh}}, \bibinfo {author}
  {\bibfnamefont {N.~D.~T.}\ \bibnamefont {Kien}},\ and\ \bibinfo {author}
  {\bibfnamefont {D.~X.}\ \bibnamefont {Viet}},\ }\bibfield  {title} {\bibinfo
  {title} {Binder ratio in the two-dimensional q-state clock model},\
  }\href@noop {} {\bibfield  {journal} {\bibinfo  {journal} {Physical Review
  E}\ }\textbf {\bibinfo {volume} {106}},\ \bibinfo {pages} {034138} (\bibinfo
  {year} {2022})}\BibitemShut {NoStop}%
\bibitem [{\citenamefont {Popov}\ \emph {et~al.}(2019)\citenamefont {Popov},
  \citenamefont {Popova},\ and\ \citenamefont {Prudnikov}}]{Popov.2019}%
  \BibitemOpen
  \bibfield  {author} {\bibinfo {author} {\bibfnamefont {I.~S.}\ \bibnamefont
  {Popov}}, \bibinfo {author} {\bibfnamefont {A.~P.}\ \bibnamefont {Popova}},\
  and\ \bibinfo {author} {\bibfnamefont {P.~V.}\ \bibnamefont {Prudnikov}},\
  }\bibfield  {title} {\bibinfo {title} {{Non-equilibrium vortex annealing of
  structural disorder in Berezinskii-Kosterlitz-Thouless dynamics of the
  two-dimensional XY-model}},\ }\href
  {https://doi.org/10.1209/0295-5075/128/26002} {\bibfield  {journal} {\bibinfo
   {journal} {EPL}\ }\textbf {\bibinfo {volume} {128}},\ \bibinfo {pages}
  {26002} (\bibinfo {year} {2019})}\BibitemShut {NoStop}%
\bibitem [{\citenamefont {Solon}\ and\ \citenamefont
  {Tailleur}(2013)}]{solon2013revisiting}%
  \BibitemOpen
  \bibfield  {author} {\bibinfo {author} {\bibfnamefont {A.}~\bibnamefont
  {Solon}}\ and\ \bibinfo {author} {\bibfnamefont {J.}~\bibnamefont
  {Tailleur}},\ }\bibfield  {title} {\bibinfo {title} {Revisiting the flocking
  transition using active spins},\ }\href@noop {} {\bibfield  {journal}
  {\bibinfo  {journal} {Physical Review Letters}\ }\textbf {\bibinfo {volume}
  {111}},\ \bibinfo {pages} {078101} (\bibinfo {year} {2013})}\BibitemShut
  {NoStop}%
\bibitem [{\citenamefont {Solon}\ \emph {et~al.}(2015)\citenamefont {Solon},
  \citenamefont {Chat{\'e}},\ and\ \citenamefont {Tailleur}}]{solon2015phase}%
  \BibitemOpen
  \bibfield  {author} {\bibinfo {author} {\bibfnamefont {A.~P.}\ \bibnamefont
  {Solon}}, \bibinfo {author} {\bibfnamefont {H.}~\bibnamefont {Chat{\'e}}},\
  and\ \bibinfo {author} {\bibfnamefont {J.}~\bibnamefont {Tailleur}},\
  }\bibfield  {title} {\bibinfo {title} {From phase to microphase separation in
  flocking models: The essential role of nonequilibrium fluctuations},\
  }\href@noop {} {\bibfield  {journal} {\bibinfo  {journal} {Physical Review
  Letters}\ }\textbf {\bibinfo {volume} {114}},\ \bibinfo {pages} {068101}
  (\bibinfo {year} {2015})}\BibitemShut {NoStop}%
\bibitem [{\citenamefont {Solon}\ and\ \citenamefont
  {Tailleur}(2015)}]{solon2015flocking}%
  \BibitemOpen
  \bibfield  {author} {\bibinfo {author} {\bibfnamefont {A.~P.}\ \bibnamefont
  {Solon}}\ and\ \bibinfo {author} {\bibfnamefont {J.}~\bibnamefont
  {Tailleur}},\ }\bibfield  {title} {\bibinfo {title} {Flocking with discrete
  symmetry: The two-dimensional active ising model},\ }\href@noop {} {\bibfield
   {journal} {\bibinfo  {journal} {Physical Review E}\ }\textbf {\bibinfo
  {volume} {92}},\ \bibinfo {pages} {042119} (\bibinfo {year}
  {2015})}\BibitemShut {NoStop}%
\bibitem [{\citenamefont {Chatterjee}\ \emph {et~al.}(2020)\citenamefont
  {Chatterjee}, \citenamefont {Mangeat}, \citenamefont {Paul},\ and\
  \citenamefont {Rieger}}]{chatterjee2020flocking}%
  \BibitemOpen
  \bibfield  {author} {\bibinfo {author} {\bibfnamefont {S.}~\bibnamefont
  {Chatterjee}}, \bibinfo {author} {\bibfnamefont {M.}~\bibnamefont {Mangeat}},
  \bibinfo {author} {\bibfnamefont {R.}~\bibnamefont {Paul}},\ and\ \bibinfo
  {author} {\bibfnamefont {H.}~\bibnamefont {Rieger}},\ }\bibfield  {title}
  {\bibinfo {title} {Flocking and reorientation transition in the 4-state
  active potts model},\ }\href@noop {} {\bibfield  {journal} {\bibinfo
  {journal} {EPL}\ }\textbf {\bibinfo {volume} {130}},\ \bibinfo {pages}
  {66001} (\bibinfo {year} {2020})}\BibitemShut {NoStop}%
\bibitem [{\citenamefont {Mangeat}\ \emph {et~al.}(2020)\citenamefont
  {Mangeat}, \citenamefont {Chatterjee}, \citenamefont {Paul},\ and\
  \citenamefont {Rieger}}]{mangeat2020flocking}%
  \BibitemOpen
  \bibfield  {author} {\bibinfo {author} {\bibfnamefont {M.}~\bibnamefont
  {Mangeat}}, \bibinfo {author} {\bibfnamefont {S.}~\bibnamefont {Chatterjee}},
  \bibinfo {author} {\bibfnamefont {R.}~\bibnamefont {Paul}},\ and\ \bibinfo
  {author} {\bibfnamefont {H.}~\bibnamefont {Rieger}},\ }\bibfield  {title}
  {\bibinfo {title} {Flocking with a q-fold discrete symmetry: Band-to-lane
  transition in the active potts model},\ }\href@noop {} {\bibfield  {journal}
  {\bibinfo  {journal} {Physical Review E}\ }\textbf {\bibinfo {volume}
  {102}},\ \bibinfo {pages} {042601} (\bibinfo {year} {2020})}\BibitemShut
  {NoStop}%
\bibitem [{\citenamefont {Chatterjee}\ \emph {et~al.}(2022)\citenamefont
  {Chatterjee}, \citenamefont {Mangeat},\ and\ \citenamefont
  {Rieger}}]{chatterjee2022polar}%
  \BibitemOpen
  \bibfield  {author} {\bibinfo {author} {\bibfnamefont {S.}~\bibnamefont
  {Chatterjee}}, \bibinfo {author} {\bibfnamefont {M.}~\bibnamefont
  {Mangeat}},\ and\ \bibinfo {author} {\bibfnamefont {H.}~\bibnamefont
  {Rieger}},\ }\bibfield  {title} {\bibinfo {title} {Polar flocks with
  discretized directions: The active clock model approaching the vicsek
  model},\ }\href@noop {} {\bibfield  {journal} {\bibinfo  {journal} {EPL}\
  }\textbf {\bibinfo {volume} {138}},\ \bibinfo {pages} {41001} (\bibinfo
  {year} {2022})}\BibitemShut {NoStop}%
\bibitem [{\citenamefont {Solon}\ \emph {et~al.}(2022)\citenamefont {Solon},
  \citenamefont {Chat{\'e}}, \citenamefont {Toner},\ and\ \citenamefont
  {Tailleur}}]{solon2022susceptibility}%
  \BibitemOpen
  \bibfield  {author} {\bibinfo {author} {\bibfnamefont {A.}~\bibnamefont
  {Solon}}, \bibinfo {author} {\bibfnamefont {H.}~\bibnamefont {Chat{\'e}}},
  \bibinfo {author} {\bibfnamefont {J.}~\bibnamefont {Toner}},\ and\ \bibinfo
  {author} {\bibfnamefont {J.}~\bibnamefont {Tailleur}},\ }\bibfield  {title}
  {\bibinfo {title} {Susceptibility of polar flocks to spatial anisotropy},\
  }\href@noop {} {\bibfield  {journal} {\bibinfo  {journal} {Physical Review
  Letters}\ }\textbf {\bibinfo {volume} {128}},\ \bibinfo {pages} {208004}
  (\bibinfo {year} {2022})}\BibitemShut {NoStop}%
\bibitem [{\citenamefont {Ferri}\ \emph {et~al.}(2023)\citenamefont {Ferri},
  \citenamefont {Gaya-\`Avila},\ and\ \citenamefont
  {D\'iaz-Guilera}}]{Ferri.2023}%
  \BibitemOpen
  \bibfield  {author} {\bibinfo {author} {\bibfnamefont {I.}~\bibnamefont
  {Ferri}}, \bibinfo {author} {\bibfnamefont {A.}~\bibnamefont
  {Gaya-\`Avila}},\ and\ \bibinfo {author} {\bibfnamefont {A.}~\bibnamefont
  {D\'iaz-Guilera}},\ }\bibfield  {title} {\bibinfo {title} {{Three-state
  opinion model with mobile agents}},\ }\href
  {https://doi.org/10.1063/5.0152674} {\bibfield  {journal} {\bibinfo
  {journal} {Chaos}\ }\textbf {\bibinfo {volume} {33}},\ \bibinfo {pages}
  {093121} (\bibinfo {year} {2023})}\BibitemShut {NoStop}%
\bibitem [{\citenamefont {Menzel}(2012)}]{Menzel.2012}%
  \BibitemOpen
  \bibfield  {author} {\bibinfo {author} {\bibfnamefont {A.~M.}\ \bibnamefont
  {Menzel}},\ }\bibfield  {title} {\bibinfo {title} {{Collective motion of
  binary self-propelled particle mixtures}},\ }\href
  {https://doi.org/10.1103/physreve.85.021912} {\bibfield  {journal} {\bibinfo
  {journal} {Physical Review E}\ }\textbf {\bibinfo {volume} {85}},\ \bibinfo
  {pages} {021912} (\bibinfo {year} {2012})}\BibitemShut {NoStop}%
\bibitem [{\citenamefont {Chatterjee}\ \emph {et~al.}(2023)\citenamefont
  {Chatterjee}, \citenamefont {Mangeat}, \citenamefont {Woo}, \citenamefont
  {Rieger},\ and\ \citenamefont {Noh}}]{Chatterjee.2023}%
  \BibitemOpen
  \bibfield  {author} {\bibinfo {author} {\bibfnamefont {S.}~\bibnamefont
  {Chatterjee}}, \bibinfo {author} {\bibfnamefont {M.}~\bibnamefont {Mangeat}},
  \bibinfo {author} {\bibfnamefont {C.-U.}\ \bibnamefont {Woo}}, \bibinfo
  {author} {\bibfnamefont {H.}~\bibnamefont {Rieger}},\ and\ \bibinfo {author}
  {\bibfnamefont {J.~D.}\ \bibnamefont {Noh}},\ }\bibfield  {title} {\bibinfo
  {title} {{Flocking of two unfriendly species: The two-species Vicsek
  model}},\ }\href {https://doi.org/10.1103/physreve.107.024607} {\bibfield
  {journal} {\bibinfo  {journal} {Physical Review E}\ }\textbf {\bibinfo
  {volume} {107}},\ \bibinfo {pages} {024607} (\bibinfo {year}
  {2023})}\BibitemShut {NoStop}%
\bibitem [{\citenamefont {Tobochnik}\ and\ \citenamefont
  {Chester}(1979)}]{tobochnik1979monte}%
  \BibitemOpen
  \bibfield  {author} {\bibinfo {author} {\bibfnamefont {J.}~\bibnamefont
  {Tobochnik}}\ and\ \bibinfo {author} {\bibfnamefont {G.}~\bibnamefont
  {Chester}},\ }\bibfield  {title} {\bibinfo {title} {Monte carlo study of the
  planar spin model},\ }\href@noop {} {\bibfield  {journal} {\bibinfo
  {journal} {Physical Review B}\ }\textbf {\bibinfo {volume} {20}},\ \bibinfo
  {pages} {3761} (\bibinfo {year} {1979})}\BibitemShut {NoStop}%
\bibitem [{\citenamefont {Loft}\ and\ \citenamefont
  {DeGrand}(1987)}]{loft1987numerical}%
  \BibitemOpen
  \bibfield  {author} {\bibinfo {author} {\bibfnamefont {R.}~\bibnamefont
  {Loft}}\ and\ \bibinfo {author} {\bibfnamefont {T.~A.}\ \bibnamefont
  {DeGrand}},\ }\bibfield  {title} {\bibinfo {title} {Numerical simulation of
  dynamics in the xy model},\ }\href@noop {} {\bibfield  {journal} {\bibinfo
  {journal} {Physical Review B}\ }\textbf {\bibinfo {volume} {35}},\ \bibinfo
  {pages} {8528} (\bibinfo {year} {1987})}\BibitemShut {NoStop}%
\bibitem [{\citenamefont {Goldenfeld}(1992)}]{Goldenfeld.1992}%
  \BibitemOpen
  \bibfield  {author} {\bibinfo {author} {\bibfnamefont {N.}~\bibnamefont
  {Goldenfeld}},\ }\href@noop {} {\emph {\bibinfo {title} {{Lectures on Phase
  Transitions and the Renormalization Group}}}}\ (\bibinfo  {publisher}
  {Addison-Wesley},\ \bibinfo {address} {Boston},\ \bibinfo {year}
  {1992})\BibitemShut {NoStop}%
\end{thebibliography}%

\end{document}